%% file: gauss_proc_tomo.tex
\documentclass[pra,12pt,onecolumn,amsmath,amssymb,superscriptaddress,eqsecnum,notitlepage]{revtex4-1}
\usepackage{graphicx,amsmath,relsize,epstopdf,upgreek,color,mathtools,bm,mathptmx,rotating}
\usepackage[hyphenbreaks]{breakurl}
\usepackage[colorlinks=true,linkcolor=blue,citecolor=blue,urlcolor =blue]{hyperref}

\newcommand{\ket}[1]{|{#1}\rangle}
\newcommand{\bra}[1]{\langle{#1}|}
\newcommand{\sket}[1]{|{#1}\rangle\!\rangle}
\newcommand{\sbra}[1]{\langle\!\langle{#1}|}
\newcommand{\sinner}[2]{\langle\!\langle{#1}|{#2}\rangle\!\rangle}
\newcommand{\opinner}[3]{\langle{#1}|{#2}|{#3}\rangle}
\newcommand{\inner}[2]{\langle{#1}|{#2}\rangle}
\newcommand{\rvec}[1]{\pmb{#1}}
\newcommand{\dyadic}[1]{\pmb{#1}}
\newcommand{\tr}[1]{\mathrm{tr}\!\left\{#1\right\}}
\newcommand{\D}{\mathrm{d}}
\newcommand{\I}{\mathrm{i}}
\newcommand{\TP}[1]{{#1}^\top}
\newcommand{\E}[1]{\mathrm{e}^{\mbox{\footnotesize$#1$}}}
\newcommand{\Tr}[1]{\mathrm{Tr}\!\left\{#1\right\}}

\newcommand{\DET}[1]{\det\!\left\{#1\right\}}
\newcommand{\RE}[1]{\mathrm{Re}\!\left\{#1\right\}}

\begin{document}
	
\allowdisplaybreaks

\title{Highly accurate Gaussian process tomography with geometrical sets of coherent states}

\author{Yong Siah Teo}
\email{yong.siah.teo@gmail.com}
\address{Department of Physics and Astronomy, Seoul National University, 08826 Seoul, Korea}
\author{Kimin Park}
\address{Department of Optics, Palack{\'y} University, 17. listopadu 12, 77146 Olomouc, Czech Republic}
\author{Seongwook Shin}
\address{Department of Physics and Astronomy, Seoul National University, 08826 Seoul, Korea}
\author{Hyunseok Jeong}
\address{Department of Physics and Astronomy, Seoul National University, 08826 Seoul, Korea}
\author{Petr Marek}
\address{Department of Optics, Palack{\'y} University, 17. listopadu 12, 77146 Olomouc, Czech Republic}

\begin{abstract}
  We propose a practical strategy for choosing sets of input coherent states that are near-optimal for reconstructing single-mode Gaussian quantum processes with output-state heterodyne measurements. We first derive analytical expressions for the mean squared-error that quantifies the reconstruction accuracy for general process tomography and large data. Using such expressions, upon relaxing the trace-preserving constraint, we introduce an error-reducing set of input coherent states that is independent of the measurement data or the unknown true process---the geometrical set. We numerically show that process reconstruction from such input coherent states is nearly as accurate as that from the best possible set of coherent states chosen with the complete knowledge about the process. This allows us to efficiently characterize Gaussian processes even with reasonably low-energy coherent states. We numerically observe that the geometrical strategy without trace preservation beats all nonadaptive strategies for arbitrary trace-preserving Gaussian processes of typical parameter ranges so long as the displacement components are not too large.
\end{abstract}


\maketitle

\section{Introduction}

Continuous-variable~(CV) systems play an important role in quantum information theory~\cite{Braunstein:2005aa,Ferraro:2005ns,CV2007:aa,Andersen:2010ng,Adesso:2014pm,Ruppert:2014aa,Ruppert:2019aa}. Gaussian states~\cite{Ferraro:2005ns}, for example, form the basic ingredients in key discussions of CV quantum information processing, notably for the study of secure quantum key distribution protocols~\cite{Lorenz:2004aa, Lance:2005aa,Scarani:2009cq,Weedbrook:2012ag}. These are quantum states described by Gaussian quasiprobability distributions~\cite{Wigner:1932aa,Cahill:1969qd}, which include the set of squeezed coherent states. The primary engines that generate these states are Gaussian processes, which are quantum processes that are also representable by a Gaussian quasidistribution. Gaussian quantum processes have been widely studied, especially in the context of channel capacity and quantum communication~\cite{Holevo:1999aa,Eisert:2007aa,Holevo:2007aa,Smith:2011aa,Lupo:2011aa,Holevo:2012aa,Siudzinska:2019aa}. 

Proper characterization of Gaussian quantum processes is crucial to ensure that Gaussian resources are reliably generated and utilized. Techniques in multiparameter estimation are commonly well-sought tools for this purposes, but very often, they are used to primarily investigate the quantum Fisher information~\cite{Braunstein:1994aa,Gross:2020aa,Kull:2020aa,DD:2020aa,Safranek:2016aa,Rosanna:2018aa} that bounds the mean squared-error of the estimated parameters. This requires optimal output-state measurements that are technically challenging to achieve in practice~\cite{Oh:2019aa}.

In this work, we shall explore a highly feasible route to optimal Gaussian process tomography that is much more accessible in experiments using coherent input states~\cite{Rahimi-Keshari:2011aa} that can be readily prepared with a well-controlled laser source. To this end, we search for a computationally efficient set of input states that lead to near-optimal precision given a fixed measurement acting on the output states. We shall consider heterodyne detection~\cite{Arthurs:1965al,Yuen:1982hh,Arthurs:1988aa,Martens:1990al,Martens:1991aa,Raymer:1994aj,Trifonov:2001up,Werner:2004as} as the output-state measurement for the exclusive advantage of its tomographic performance in reconstructing Gaussian states~\cite{Rehacek:2015qp,Muller:2016da,Teo:2017aa} over homodyne detection~\cite{Yuen:1983ba,Abbas:1983ak, Schumaker:1984qm}, both of which essentially constitute the typical CV measurements that can be carried out in practice. Another key departure from previous work is that generic Gaussian processes shall be considered in our study, rather than just their subclasses.

The mean squared-error~(MSE) for all the parameters characterizing the unknown Gaussian process is adopted as the figure of merit for the reconstruction quality. To analyze the MSE for general Gaussian processes with large data samples, we shall derive its asymptotic formulas by extending methods previously developed for quantum states~\cite{Teo:2017aa,Zhu:2014aa,Teo:2015qs}. Next, without imposing the trace-preserving~(TP) constraint, we construct a convenient set of input coherent states that minimize the MSE Cauchy--Schwarz \emph{upper bound} for the unknown Gaussian process. We demonstrate that such states give an MSE that is almost identical to the optimal value provided by the best nonadaptive set of input states obtainable only with the complete knowledge about the process of interest. This near-optimality turns even reasonably low-energy coherent states into formidable resources for reconstructing Gaussian processes. Such an input set is ``geometrical'' since the phase-space arrangement of these coherent states is predetermined by only the output-state measurements employed and nothing else. Furthermore, we show numerically that for \emph{arbitrary} completely-positive-trace-preserving~(CPTP) Gaussian processes of parameter ranges typically considered in experiments (to be specified more concretely in Sec.~\ref{sec:results}), the non-TP geometrical strategy emerges as the optimal nonadaptive strategy by asymptotically outperforming the best TP strategy so long as the process displacement components are not very large.

After some background introduction to the general formalism of Gaussian processes in Sec.~\ref{sec:bkgd}, Sec.~\ref{sec:mse} shall be devoted to the explanation and derivation of the MSE formulas for both TP and non-TP reconstruction methods. With the aid of these formulas, Sec.~\ref{sec:geom} then proceeds with the construction of geometrical input states. Finally, Sec.~\ref{sec:results} compares the geometrical strategy with existing common nonadaptive input-state strategies for realistic CPTP Gaussian processes.

\section{Characterization of Gaussian processes}
\label{sec:bkgd}

A physical quantum process $\Phi$ transforms an input state $\rho_\textsc{in}$ into the output state $\rho_\textsc{out}=\Phi[\rho_\textsc{in}]$. A standard operational description for the quantum process $\Phi$ makes use of the \emph{Choi-Jamio\l kowski} formalism, which essentially states that all information about $\Phi$ is encoded into a positive operator ($\rho_\Phi$). Additionally, we say that $\Phi$ is Gaussian if it possesses a two-mode Gaussian quasidistribution. In this case, it is convenient to represent $\Phi$ by its Husimi~Q function
\begin{equation}
Q_\Phi=\exp\,(-\rvec{Z}^\dag \dyadic{A}\, \rvec{Z}+\rvec{B}^\dag\rvec{Z}+c_0)\,,
\end{equation} 
which is defined by a complex matrix $\dyadic{A}$, a complex column $\rvec{B}$ (both in the computational basis) and a real constant $c_0$. Here $\rvec{Z}=\TP{(\rvec{z}_1\,\,\rvec{z}_2)}$ consolidates the complex variables labeling the process input~[$\rvec{z}_1=\TP{(z_1\,\,z_1^*)}$] and output~[$\rvec{z}_2=\TP{(z_2\,\,z_2^*)}$] modes. 

\emph{Gaussian process tomography} pertains to the characterization of any given unknown $\rho_\Phi$ on the premise that its $Q_\Phi$ is Gaussian. The connection between $\rho_\Phi$ and $Q_\Phi$ is made by \emph{heterodyne measurements}~\cite{Arthurs:1965al,Yuen:1982hh,Arthurs:1988aa,Martens:1990al,Martens:1991aa,Raymer:1994aj,Trifonov:2001up,Werner:2004as} that sample the overcomplete set of coherent states $\{\ket{z}\bra{z}\}$ to probe the output state $\rho_\textsc{out}=\mathrm{tr}_1\{\TP{\rho}_\textsc{in}\otimes 1\, \rho_\Phi\}$, where the transposition is defined for the Fock basis in which all matrices are written in this article. Apart from directly recovering the Q-function parameters, these measurements are also known to give a smaller MSE for characterizing covariance matrices of Gaussian and broad classes of non-Gaussian quantum states compared to its homodyne counterpart~\cite{Rehacek:2015qp,Muller:2016da,Teo:2017aa}. Another reason for this choice of measurements is that when coherent input states $\{\ket{\alpha}\bra{\alpha}\}$ are used, the heterodyne measurement is equivalent to a direct sampling of the process Q~function, as $Q_\Phi(\alpha,\alpha^*,z,z^*)=\opinner{\alpha}{\rho_\textsc{out}}{\alpha}$.

For a completely-positive (CP) $\Phi$ ($\rho_\Phi\geq0$), $\dyadic{A}$, $\rvec{B}$ and $c_0$ are constrained such that $Q_\Phi$ is positive and square-integrable. One way to identify these constraints systematically is by reverting to the real phase-space representation: $Q_\Phi=\exp\,(-\TP{\rvec{R}} \dyadic{A}'\, \rvec{R}+\TP{\rvec{B}'}\rvec{R}+c_0)$ with $\rvec{R}=\TP{(x_1\,\,p_1\,\,x_2\,\,p_2)}$. This is done by recognizing that the transformations $\rvec{Z}=\dyadic{U}\,\rvec{R}$, $\dyadic{A}'=\dyadic{U}^\dag\dyadic{A}\,\dyadic{U}$ and $\rvec{B}'=\dyadic{U}^\dag\,\rvec{B}$ are exacted with the unitary matrix $\dyadic{U}=\dyadic{1}\otimes \dyadic{U}_0$ and $\dyadic{U}_0=\begin{pmatrix}1 & \I\\1 &-\I\end{pmatrix}/\sqrt{2}$. It is now clear that the conditions $\dyadic{A}'\geq0$ and $\dyadic{A}\geq0$ are necessary for $Q_\Phi$ to be real and square-integrable. These give a total of 15 independent real parameters, that is 10 from $\dyadic{A}$, 4 from $\rvec{B}$, and $c_0$. We may parametrize $\dyadic{A}$ and $\rvec{B}$ as
\begin{align}
\dyadic{A}=&\,\begin{pmatrix}
\dyadic{A}_1 & \dyadic{A}_2\\
\dyadic{A}^\dag_2 & \dyadic{A}_3
\end{pmatrix}\,,\,\,\dyadic{A}_1=\begin{pmatrix}
a_1/2 & -c_1^*\\
-c_1 & a_1/2
\end{pmatrix}\,,\nonumber\\
\dyadic{A}_2=&\,\dfrac{1}{2}\begin{pmatrix}
g_2 & g_1^*\\
g_1 & g_2^*
\end{pmatrix}\,,\,\,\dyadic{A}_3=\begin{pmatrix}
a_2/2 & -c_2^*\\
-c_2 & a_2/2
\end{pmatrix}\,,\nonumber\\
\rvec{B}=&\,\begin{pmatrix}
\rvec{b}_1\\
\rvec{b}_2
\end{pmatrix}\,,\,\,\rvec{b}_1=\begin{pmatrix}
b_1\\
b_1^*
\end{pmatrix}\,,\,\,\rvec{b}_2=\begin{pmatrix}
b_2\\
b_2^*
\end{pmatrix}\,.
\label{eq:Gauss_param}
\end{align}

A useful $\Phi$ in quantum information theory is typically also TP $(\tr{\rho_\textsc{in}}=1=\tr{\rho_\textsc{out}})$. Under this constraint, for an invertible $\dyadic{A}_3$, it is shown in Appendix~\ref{app:TP} that 6 of the 15 real parameters are fixed by the rest inasmuch as
\begin{align}
\dyadic{A}_1=&\,\,\dyadic{A}_2\,\dyadic{A}_3^{-1}\,\dyadic{A}_2^\dag\qquad\qquad\qquad\qquad\,\,\,\,\,\, (\text{3 parameters})\,,\nonumber\\
\rvec{b}_1=&\,\,\dyadic{A}_2\,\dyadic{A}_3^{-1}\rvec{b}_2\qquad\qquad\qquad\qquad\,\,\,\,\,\,\,\, (\text{2 parameters})\,,\nonumber\\
c_0=&\,\log(2\sqrt{\DET{\dyadic{A}_3}})-\dfrac{1}{4}\rvec{b}^\dag_2\,\dyadic{A}_3^{-1}\rvec{b}_2\quad\! (\text{1 parameter})\,.
\label{eq:TP_constr}
\end{align} 
As a simple example, if we consider a beam splitter that transforms a pair of input mode operators $a$ and $b$ into the pair of output operators $c=a\cos\theta+b\sin\theta$ and $d=-a\sin\theta+b\cos\theta$, then the relevant Choi-Jamio{\l}kowski operator for \emph{a single output mode} ($c$) clearly describes a CPTP process and possesses the Q~function $Q_\Phi=\exp(-|z_1|^2(\cos\theta)^2-|z_2|^2+z_1z_2\cos\theta+z^*_1z^*_2\cos\theta)$~\cite{Wang:2013aa}. In this case, a consistency check gives $\rvec{b}_2=\rvec{0}$, $\dyadic{A}_3=\dyadic{1}/2$, $\dyadic{A}_2=-(\cos\theta)\,\dyadic{\sigma}_x/2$, $\dyadic{A}_1=\dyadic{A}_2\,\dyadic{A}_3^{-1}\dyadic{A}_2^\dag=(\cos\theta)^2\,\dyadic{1}/2$, $\rvec{b}_1=\rvec{0}$ and $c_0=0$, with $\dyadic{\sigma}_x$ being the usual Pauli $x$ matrix in the standard basis.

If the TP constraint is absent from the process reconstruction, the parameter $c_0$ is not estimable since any experimental data can only recover $\rho_\Phi$ uniquely up to a constant multiple~\cite{Bongioanni:2010aa,Teo:2020aa}, such that $\Phi$ may only be fully characterized up to its operator trace. Therefore, the complete characterization of a general single-mode Gaussian $\Phi$ requires 14 real recoverable parameters:
\begin{equation}
\rvec{x}=\TP{(a_1\,\,a_2\,\,b_{1,\mathrm{r}}\,\,b_{1,\mathrm{i}}\,\,b_{2,\mathrm{r}}\,\,b_{2,\mathrm{i}}\,\,c_{1,\mathrm{r}}\,\,c_{1,\mathrm{i}}\,\,c_{2,\mathrm{r}}\,\,c_{2,\mathrm{i}}\,\,g_{1,\mathrm{r}}\,\,g_{1,\mathrm{i}}\,\,g_{2,\mathrm{r}}\,\,g_{2,\mathrm{i}})}\,,
\label{eq:param_nonTP}
\end{equation}
where the subscripts r and i denote the real and imaginary parts of a complex parameter. Formally, the coherent-state sampling measurements by heterodyning gather raw data sampled from the Q~function $Q_\textsc{out}(\rvec{x};z_2\equiv z,z^*_2\equiv z^*)$ of $\rho_\textsc{out}$ (originating from a given $\rho_\textsc{in}$) that encodes $\rvec{x}$. Numerical techniques are then used to obtain the estimator $\widehat{\rvec{x}}$ (distinguished from the true parameter by a caret).

In this work, we focus on studying the accuracy of $\widehat{\rvec{x}}$ for a given unknown $\rvec{x}$. This may be quantified by the MSE~$\overline{(\widehat{\rvec{x}}-\rvec{x})^2}$, where the overline denotes an average over all possible data of a fixed total sample size. For an analytical study, we shall investigate the asymptotic expression of the MSE that applies to typical tomography situations involving large datasets.

\begin{figure}[t]
	\centering
	\includegraphics[width=1\columnwidth]{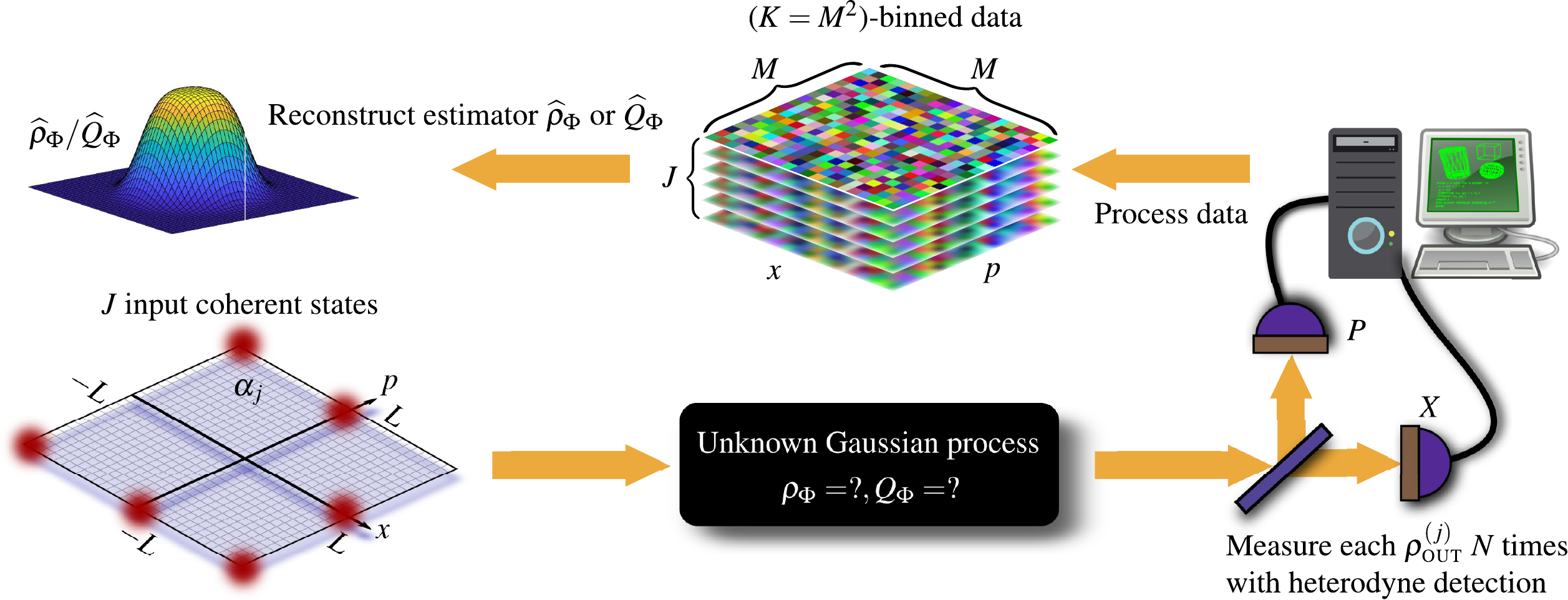}
	\caption{\label{fig:gauss_opt_tomo}Schematic diagram of (Gaussian) process characterization. A set of $J$ input coherent states of amplitudes $\{\alpha_j\}$, lying in the phase-space box region $-L\leq\alpha_{j,\mathrm{r}},\alpha_{j,\mathrm{i}}\leq L$, are fed to the unknown Gaussian process $\Phi$. Heterodyne measurements, implemented through simultaneous measurements of the position ($X$) and momentum ($P$) quadratures, are performed $N$ times on each of the respective output states. The collected data from all $J$ output states are processed on the discretized phase space of $K=M^2$ bins, after which the Q-function estimator $\widehat{Q}_\Phi$ is reconstructed from the binned data.}
\end{figure}

\section{Mean squared-error formulas}
\label{sec:mse}

\subsection{Relaxation of the TP constraint}
\label{subsec:mse_nonTP}

We first investigate the case where coherent states are used as input states for characterizing an unknown, generally non-TP Gaussian $\Phi$. Given a $\rho_\textsc{in}=\ket{\alpha}\bra{\alpha}$, the output Gaussian Q-function~$Q_\textsc{out}$ can be written in the form
\begin{align}
Q_\textsc{out}=&\,\E{-\TP{\rvec{v}}\rvec{x}'}\,,\nonumber\\
\frac{\TP{\rvec{v}}}{2}=&\,\Big(\frac{|\alpha|^2}{2}\,\,\,\,\frac{|z|^2}{2}\,\,\,\,\alpha_\mathrm{r}\,\,\,\,\alpha_\mathrm{i}\,\,\,\,z_\mathrm{r}\,\,\,\,-z_\mathrm{i}\,\,\,\,(\alpha^2)_\mathrm{r}\,\,\,\,(\alpha^2)_\mathrm{i}\,\,\,\,(z^2)_\mathrm{r}\,\,\,\,-(z^2)_\mathrm{i}\nonumber\\
&\,\,\,\,\,(\alpha z^*)_\mathrm{r}\,\,\,\,(\alpha z^*)_\mathrm{i}\,\,\,\,(\alpha^*z^*)_\mathrm{r}\,\,\,\,(\alpha^*z^*)_\mathrm{i}\,\,\,\,\frac{1}{2}\Big)\,,
\label{eq:Qout_coh}
\end{align}
where $\rvec{x}'=\TP{(\rvec{x}\,\,\,c_0)}$ carries the non-estimable $c_0$. In Gaussian process tomography, $\rvec{x}$ may be extracted from a discretized system of equations governed by \eqref{eq:Qout_coh}. The latter is established by first sending $J$ input coherent states $\{\ket{\alpha_j}\bra{\alpha_j}\}^J_{j=1}$, next performing heterodyne measurements on all corresponding output states $\rho^{(j)}_\textsc{out}$, and later bin the collected data into an $M\times M$ phase-space grid. If the number of phase-space bins $K=M^2$ is large, the final reconstructed $\widehat{\rvec{x}}$ ($\widehat{Q}_\textsc{out}$) should approximate the actual $\rvec{x}$ ($Q_\textsc{out}$) efficiently. The entire flow of Gaussian-process characterization is concisely pictorialized in Fig.~\ref{fig:gauss_opt_tomo}.

For a sufficiently large $J$, we can extract $\rvec{x}=\widetilde{\dyadic{V}^-}\rvec{u}$ by inverting the exponent of Eq.~\eqref{eq:Qout_coh} after taking the logarithm on both sides---the \emph{logarithmic inversion}~(LI) procedure. Here $\widetilde{\dyadic{V}^-}$ is the matrix of the first 14 rows of the \emph{left-pseudoinverse} $\dyadic{V}^-$ $(\dyadic{V}^-\dyadic{V}=\dyadic{1})$ that is defined for the \mbox{$JK\times15$} matrix $\dyadic{V}=(\rvec{v}_{\alpha_1,z_1}\,\,\ldots\,\,\rvec{v}_{\alpha_1,z_K}\,\,\rvec{v}_{\alpha_2,z_1}\,\,\ldots\,\,\rvec{v}_{\alpha_2,z_K}\,\,\ldots\,\,\rvec{v}_{\alpha_J,z_1}\,\,\rvec{v}_{\alpha_J,z_K})^\textsc{t}$ and \mbox{$JK\times1$} column $\rvec{u}=(-\log p_{11}\,\,\ldots\,\, -\log p_{1K}\,\, -\log p_{21}\,\,\ldots\,\, -\log p_{2K}\,\,\ldots\,\, -\log p_{J1}\,\,\ldots\,\, -\log p_{JK})^\textsc{t}$ acquired from an \mbox{$M\times M$} phase-space grid. Each probability $p_{jk}$ is proportional to $Q^{(j)}_\textsc{out}(\rvec{x};z_k,z^*_k)$ up to proper normalization as a consequence of binning.

For LI to be successful, the system $\rvec{u}=\dyadic{V}\rvec{x}'$ must be informationally complete~(IC), that is, there exists a $\dyadic{V}^-$ that is uniquely given by $\dyadic{V}^{-}=(\dyadic{V}^\dag\dyadic{V})^{-1}\dyadic{V}^\dag$. This implies that measurement data collected with such a set of input states uniquely characterize the unknown Gaussian process. In equivalent linear-algebraic terms, an IC set of linearly independent input states gives rise to an invertible \emph{Gram matrix} $\dyadic{G}=\dyadic{V}^\dag\dyadic{V}$ if $J\geq6$. To understand why this is the case, we observe that as 6 out of the 15 terms in $\rvec{v}$ do not depend on $z$, when $J<6$, there naturally exists at least one null right eigenvector $\rvec{e}$ for $\dyadic{V}$ of the form $\rvec{e}=\TP{(e_1\,\,\,0\,\,\,e_2\,\,\,e_3\,\,\,0\,\,\,0\,\,\,e_4\,\,\,e_5\,\,\,0\,\,\,0\,\,\,0\,\,\,0\,\,\,0\,\,\,0\,\,\,e_6)}$, where the 6-dimensional $(e_1\,\,\,e_2\,\,\,e_3\,\,\,e_4\,\,\,e_5\,\,\,e_6)$ is orthogonal to $(|\alpha_j|^2\,\,\,2\alpha_{j,\mathrm{r}}\,\,\,2\alpha_{j,\mathrm{i}}\,\,\,2(\alpha_j^2)_\mathrm{r}\,\,\,2(\alpha_j^2)_\mathrm{i}\,\,\,1)$ for any amplitude $\alpha_j$. This observation is therefore consistent with the alternative arguments in~\cite{Wang:2013aa}.

\begin{figure}[t]
	\centering
	\includegraphics[width=0.7\columnwidth]{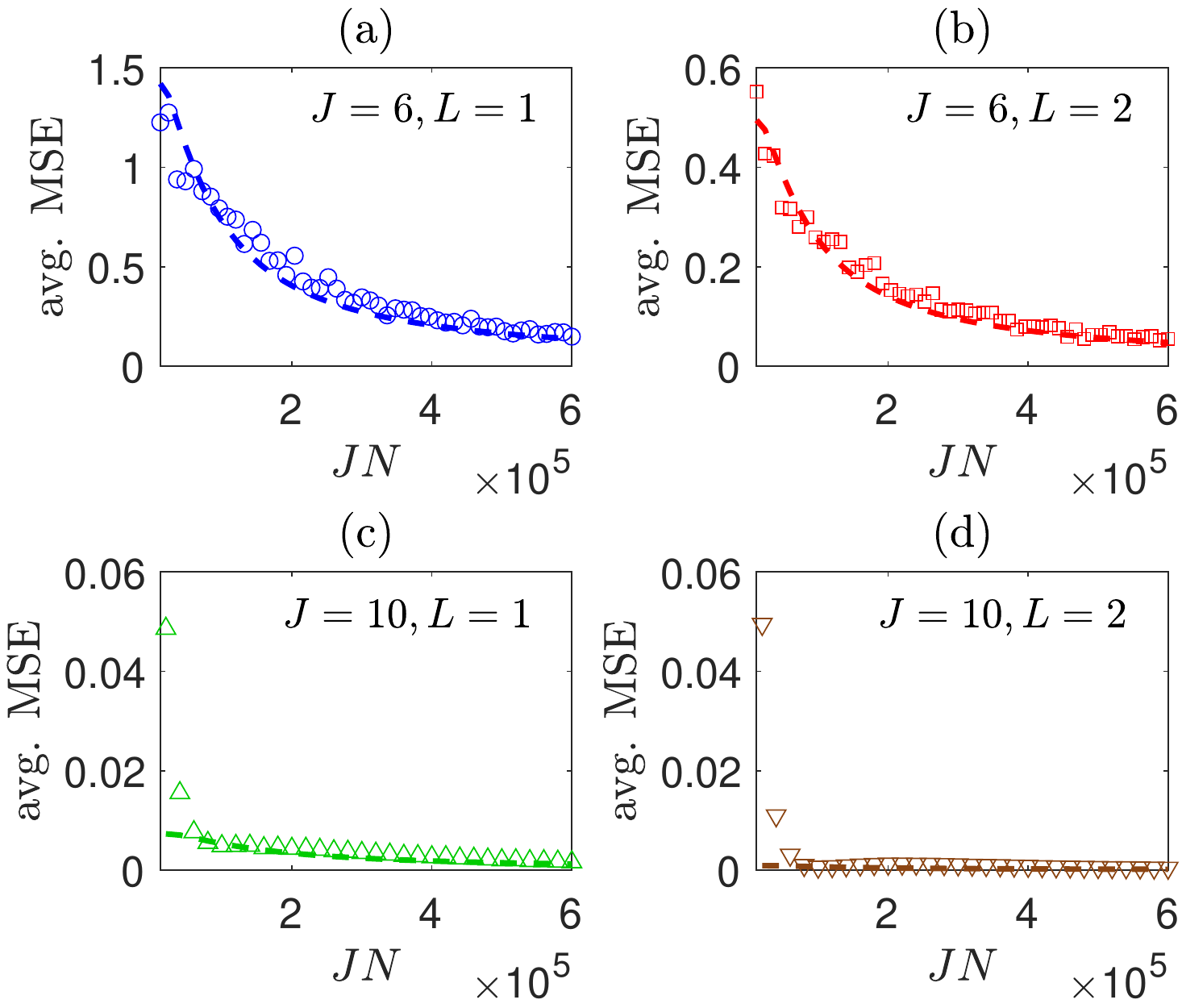}
	\caption{\label{fig:th_sim_coh}The resource performance of four simulated tomography scenarios on a random Gaussian process using various numbers of randomly-chosen input coherent states and a phase-space grid of $K=400$ as an illustration. All MSEs are computed over all the 14 estimable parameters, and averaged over 100 experiments and 50 random sets of input states for each $J$ and $N$. Dashed curves are results obtained from the analytical formula in Eq.~\eqref{eq:asymp_cohIN_hetOUT} that asymptotically approximates the MSE based on LI reconstruction. The general trend is consistent with the physical understanding that the accuracy of $\widehat{\rvec{x}}$ improves when $J$ and $L$ are large.}
\end{figure}

In realistic scenarios, the log-probability column $\rvec{u}$ is to be replaced by the column of relative log-frequencies $\widehat{\rvec{u}}=(-\log \nu_{jk})$ that reflects the physical relative photodetection counts. Note that $\sum_k\nu_{jk}=1$, and $\nu_{jk}\rightarrow p_{jk}$ as $N\gg1$ in a statistically consistent setting. As a consequence, the LI procedure that now handles these noisy data $\nu_{jk}$ should be modified. As the counts are noisy with statistical fluctuation, for any finite number of sampling copies $N$ per input state, there very likely exist entries in $\widehat{\rvec{u}}$ that are infinite ($\nu_{jk}=0$ for some $j$ and $k$), especially when the corresponding Q-function magnitudes are small. To cope with statistical noise in LI, one may consider only finite entries of $\widehat{\rvec{u}}$. After some statistical reasoning (see Appendix~\ref{app:asymp_log}), we obtain the asymptotic expression
\begin{align}
\mathrm{MSE}\equiv&\,\overline{(\widehat{\rvec{x}}-\rvec{x})^2}=\dfrac{1}{N}\mathrm{Tr}\Big\{\widetilde{\dyadic{V}^-}^\dag\widetilde{\dyadic{V}^-}\,\dyadic{Y}\Big\}\,,\nonumber\\
Y_{jk,j'k'}=&\,\delta_{j,j'}[1-(1-\widetilde{p}_{jk})^N][1-(1-\widetilde{p}_{jk'})^N]\left(\dfrac{\delta_{k,k'}}{\widetilde{p}_{jk}}-1\right)\,,
\label{eq:asymp_cohIN_hetOUT}
\end{align}
where we note that $\sum_k\widetilde{p}_{jk}=1$ are the normalized true probabilities related to $p_{jk}$ through $\widetilde{p}_{jk}=p_{jk}/\sum_kp_{jk}$. Figure~\ref{fig:th_sim_coh} illustrates the positive match between the theoretical expression in \eqref{eq:asymp_cohIN_hetOUT} and simulation results for a given Gaussian process. The real and imaginary parts of the complex input coherent-state amplitude $\alpha_j=\alpha_{j,\mathrm{r}}+\I\,\alpha_{j,\mathrm{i}}$ are chosen from the closed interval [$-L,L$], where the influence of $L$ on the characterization quality of $\Phi$ is explored. We assume that $\gamma_j\equiv\sum_kp_{jk}$ are known with small statistical fluctuation up to a scalar multiple.

We stress that the LI estimator $\widehat{x}$ introduced here, while useful as a formalism for an analytical grasp of the actual characterization problem, usually does not lead to a physical process estimator $\widehat{\Phi}$, since the inversion procedure pays no attention to the positivity requirement for the estimated complex $\widehat{\dyadic{A}}$ matrix. Numerically, it is possible to enforce such a positivity constraint in LI, in which case the resulting estimator will have some statistical bias and an MSE that deviates slightly from the expression in \eqref{eq:asymp_cohIN_hetOUT}. One may also choose to perform LI followed by a projection onto the real and positive $\dyadic{A}'$-space in the real phase-space representation, as previously discussed in Sec.~\ref{sec:bkgd}, to obtain a sufficiently good physical estimator $\widehat{x}$. Supposing that the estimated real matrix $\widehat{\dyadic{A}'}=\dyadic{U}_\text{diag}\,\dyadic{D}\,\dyadic{U}^\dag_\text{diag}$ is diagonalized by the unitary $\dyadic{U}_\text{diag}$, this projection is done through the map $\widehat{\dyadic{A}'}\mapsto\widehat{\dyadic{A}}'_\text{physical}=\mathrm{Tr}\big\{\dyadic{D}\big\}\dyadic{U}_\text{diag}\,\dyadic{D}_+\dyadic{U}^\dag_\text{diag}/\Tr{\dyadic{D}_+}$, where $\dyadic{D}_+$ is essentially the diagonal matrix $\dyadic{D}$ with all negative eigenvalues set to zero. 

While LI with positivity constraint and the projection method give highly similar estimators for sufficiently large $N$, the scaling behaviors in $JN$ for both methods generally vary. To put things on firmer statistical grounds, more meaningful estimators, such as the \emph{maximum-likelihood}~(ML) estimators, should be considered. In the context of non-TP process characterization, ML asymptotically gives very similar reconstructions to LI under the physical process constraints. Section~\ref{subsec:ML_TP} provides an explanation regarding this connection and presents a recipe for the ML reconstruction prescription.

\subsection{Imposition of the TP constraint}
\label{subsec:ML_TP}

If the unknown Gaussian process $\Phi$ is TP, then the constraints specified in \eqref{eq:TP_constr} dictate that 9 parameters are enough to characterize $\Phi$:
\begin{equation}
\rvec{x}=\TP{(a_2\,\,b_{2,\mathrm{r}}\,\,b_{2,\mathrm{i}}\,\,c_{2,\mathrm{r}}\,\,c_{2,\mathrm{i}}\,\,g_{1,\mathrm{r}}\,\,g_{1,\mathrm{i}}\,\,g_{2,\mathrm{r}}\,\,g_{2,\mathrm{i}})}\,.
\label{eq:param_TP}
\end{equation}
We note that in this case, $\dyadic{A}\geq0$ so long as $\dyadic{A}_3>0$, since we may write
\begin{equation}
\dyadic{A}\,\widehat{=}\begin{pmatrix}
\dyadic{A}_2\,\dyadic{A}_3^{-1/2}\\
\dyadic{A}_3^{1/2}
\end{pmatrix}\begin{pmatrix}
\dyadic{A}_3^{-1/2}\dyadic{A}_2^\dag\quad\dyadic{A}_3^{1/2}
\end{pmatrix}
\end{equation} 
with well-defined matrix square-roots. This also implies that $\dyadic{A}$ is rank-2 and that $a_2^2-4|c_2|^2>0$ is the only necessary and sufficient positivity condition for a CPTP Gaussian $\Phi$ as no other constraints are imposed on $\dyadic{A}_2$ and $\rvec{b}_2$.

Because of the nonlinear dependence on the 9 parameters in the exponent of the output Q~function in accordance with \eqref{eq:TP_constr}, LI is no longer applicable as it only works with a highly specific form of the output Q~function stated in \eqref{eq:Qout_coh}. Instead, statistical method is usually a more favorable option to infer $\rvec{x}$ from the collected data. A popular method is to maximize the log-likelihood function $\log \mathcal{L}=\sum_{jk}\nu_{jk}\log(p_{jk}/\sum_{j'k'}p_{j'k'})$ that takes the multinomial form when each output state is measured with $N$ sampling copies of heterodyne detection independently, subject to the positivity constraint of $\dyadic{A}_3\geq\dyadic{0}$. The log-likelihood $\log \mathcal{L}$ may in general be a nonconvex function of the TP Gaussian-process parameters, so standard numerical techniques might be needed to search for its global maximum for optimal accuracy. 

We emphasize that the ML scheme may be applied to any tomographic situation, which evidently includes the characterization of non-TP processes. In this context, with respect to the variable probabilities $p'_{jk}=\exp(-\TP{\rvec{v}_{jk}}\rvec{x}')$, we consider only those $\nu_{jk}>0$ in 
\begin{equation}
\log \mathcal{L}=-\sum_{jk}\nu_{jk}\,\TP{\rvec{v}_{jk}}\rvec{x}'-\left(\sum_{j'}\gamma_{j'}\right)\log\left(\sum_{jk}\E{-\TP{\rvec{v}_{jk}}\rvec{x}'}\right)\,,
\end{equation}
where we again assume that the $\gamma_j$s can be determined through calibration procedures up to a multiplicative constant and are not part of the statistical consideration. Maximizing $\log \mathcal{L}$ involves scaling its gradient
\begin{equation}
\frac{\updelta \log \mathcal{L}}{\updelta\rvec{x}'}=\sum_{jk}\left(-\nu_{jk}+ \dfrac{\mu\,p'_{jk}}{\sum_{j'k'}p'_{j'k'}}\right)\TP{\rvec{v}_{jk}}
\end{equation}
for the parameter $\rvec{x}'$ and $\mu=\sum_{j'}\gamma_{j'}$. If the solution to $\mu p'_{jk}/\sum_{j'k'}p'_{j'k'}=\nu_{jk}$ exists under the physical constraints of the parameter estimator $\widehat{\rvec{x}}$ for which the corresponding process estimator $\widehat{\Phi}$ remains a CP process [namely $\dyadic{A}\geq0$ as stated in \eqref{eq:Gauss_param}], then the peak of $\log \mathcal{L}$ can obviously be reached by the maximization. Under this situation, both ML and LI schemes are equivalent when $p'_{jk}=\nu_{jk}$. In the hypothetical event that $\nu_{jk}$ are completely noiseless, then this solution uniquely maximizes the log-likelihood. For finite $N$, satisfying the constraints of $\rvec{x}$ almost surely leads to $p'_{jk}\neq\nu_{jk}$. Nevertheless, for sufficiently large $N$, the MSEs obtained with both schemes are typically not too far from each other in the absence of other sources of external systematic errors. 

\begin{figure}[t]
	\centering
	\includegraphics[width=0.7\columnwidth]{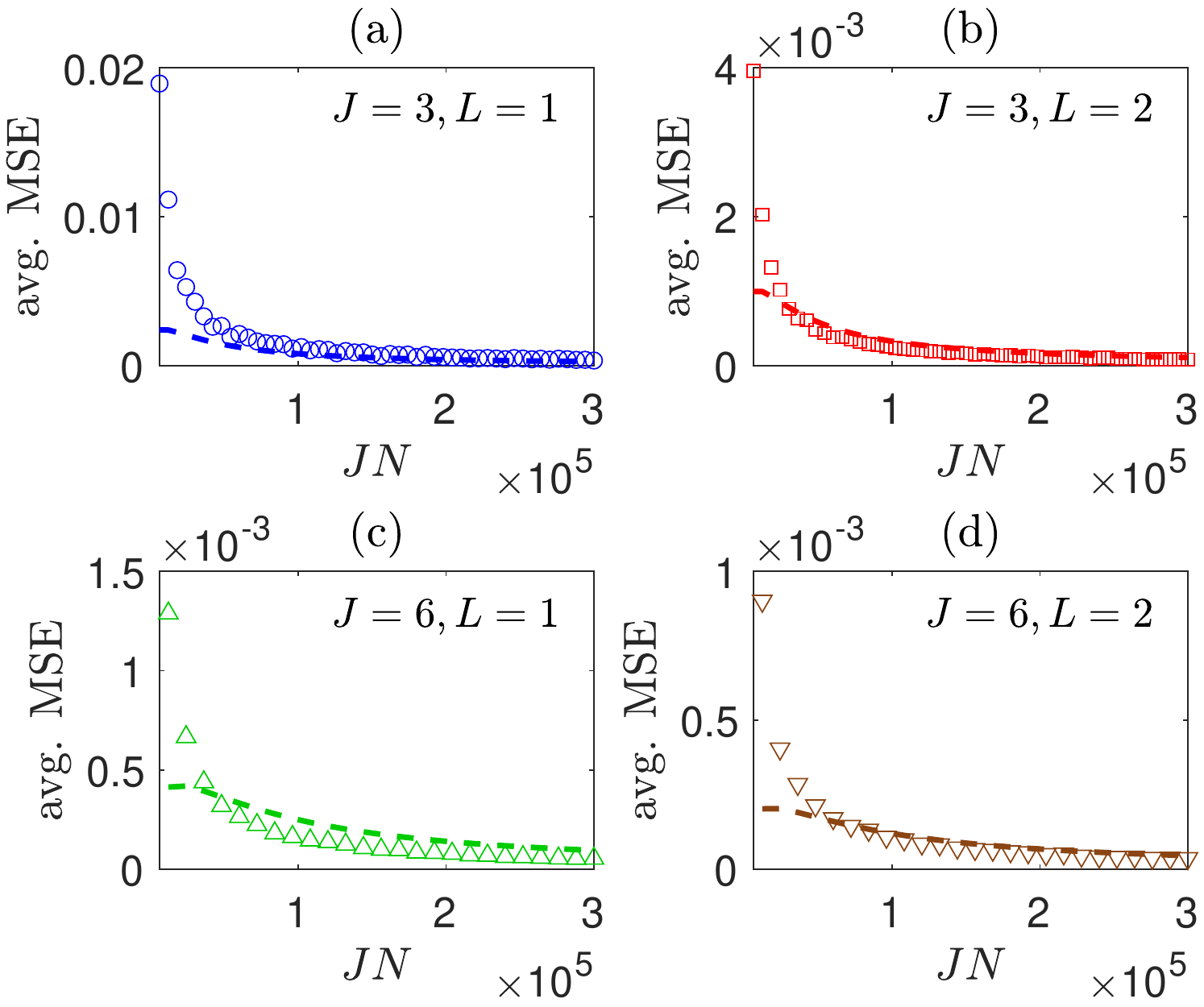}
	\caption{\label{fig:th_sim_coh_TP}The resource performance of four simulated tomography scenarios on a random Gaussian TP process of general specifications identical to those of Fig.~\ref{fig:th_sim_coh}, where 50 simulated experiments and 20 random input coherent states are used to average the MSE for the 9 CPTP parameters. Here, $J=3$ turns out to be the minimum number of input coherent states to fully determine an unknown TP Gaussian process~\cite{Wang:2013aa}. Dashed curves are results obtained from the analytical formula in Eq.~\eqref{eq:asymp_cohIN_hetOUT_TP}, which approximate the respective simulated ML MSEs well for large $JN$.}
\end{figure}

The asymptotic MSE expression for the maximum-likelihood~(ML) estimator $\widehat{\rvec{x}}_\textsc{ml}$ for $\rvec{x}$ may be approximately calculated by assuming that $N\gg1$ per output state is large enough so that $\updelta\rvec{x}\equiv\rvec{x}-\widehat{\rvec{x}}_\textsc{ml}$ is typically small. Similar to the treatment presented in Sec.~\ref{subsec:mse_nonTP}, we can define a $JK\times9$ matrix $\dyadic{V}_\textsc{tp}$ such that its $(j,k)$th row is equal to the $1\times9$ row
\begin{equation}
\TP{\rvec{v}}_\textsc{tp}=(\sbra{\dyadic{M}^\dag_{1,jk}}\dyadic{E}_1\,\,\,\,\sbra{\dyadic{M}_{2,jk}+\dyadic{M}^\dag_{2,jk}}\dyadic{E}_2\,\,\,\,\sbra{\dyadic{M}^\dag_{3,jk}}\dyadic{E}_3)\,,
\end{equation}
where the definitions of all auxiliary matrices are given in Appendix~\ref{app:asymp_log}. The principle of small variations thus states that $\dyadic{V}_\textsc{tp}\,\updelta\rvec{x}=\updelta\rvec{u}$. It follows that the asymptotic MSE for estimating the 9 independent CPTP parameters using the constrained ML method is approximately
\begin{equation}
\mathrm{MSE}\equiv\overline{(\widehat{\rvec{x}}_\textsc{ml}-\rvec{x})^2}\approx\dfrac{1}{N}\Tr{{\dyadic{V}^-_\textsc{tp}}^\dag\dyadic{V}^-_\textsc{tp}\,\dyadic{Y}}\,,
\label{eq:asymp_cohIN_hetOUT_TP}
\end{equation}
where $\dyadic{Y}$ is as specified in Eq.~\eqref{eq:asymp_cohIN_hetOUT}.

The ML estimator $\widehat{\rvec{x}}_\textsc{ml}$ is to be constrained by the positivity of $\dyadic{A}_3$ and is typically a biased estimator. In general, there could still be a small difference between the ML MSE and the right-hand side of Eq.~\eqref{eq:asymp_cohIN_hetOUT_TP} that is obtained from operator derivatives that assume the existence of open sets without parameter boundary constraints. Barring this technical issue, Fig.~\ref{fig:th_sim_coh_TP} shows that \eqref{eq:asymp_cohIN_hetOUT_TP} can still serve as a pretty good estimate for the actual MSE.

\section{Geometrical set of input coherent states}
\label{sec:geom}

The LI procedure discussed in Sec.~\ref{subsec:mse_nonTP} hinges on the existence of $\dyadic{V}^{-}$. This is again synonymous to having a $\dyadic{V}$ with no null right eigenvectors, thereby ruling out sets of coherent states with identical amplitudes $|\alpha_j|\equiv|\alpha|$ as candidates for LI, since they result in at least one such null eigenvector, namely $\dyadic{e}\propto\TP{(-1/|\alpha|^2\,\,\,0\,\,\,\ldots\,\,\,0\,\,\,1)}$, for any $J$. Therefore, choices that include the set of coherent states with complex amplitudes that form a ring of radius $r$ in phase space---$\alpha_j=r\,\E{2\pi\I\,j/J}$ for $r>0$ and $1\leq j\leq J$---are non-IC and shall result in failures of the LI scheme. In the regime of ML estimation, these symmetric sets of coherent states give a convex set of estimated parameters that are consistent with the ML probabilities obtained from the measurement data. One therefore cannot obtain a unique parameter reconstruction with such input states.

A general observation from Figs.~\ref{fig:th_sim_coh} and \ref{fig:th_sim_coh_TP} is that input coherent states with lower phase-space energies ($L=1$ for instance) also tend to give larger average MSE values. This is because a set of low-energy coherent states are typically quite closely packed in phase space and their Gram matrix $\dyadic{G}$ can at times be ill-conditioned, that is, its smallest eigenvalue can be very close to zero. Colloquially, low-energy coherent states are not very linearly independent. However, with appropriate optimization strategies, low-energy input coherent states can still be highly effective in characterizing any unknown Gaussian process with significantly higher accuracy than using random low-energy input states, thereby allowing us to maximally utilize these energy-efficient resources.

\begin{figure}[t]
	\centering
	\includegraphics[width=0.6\columnwidth]{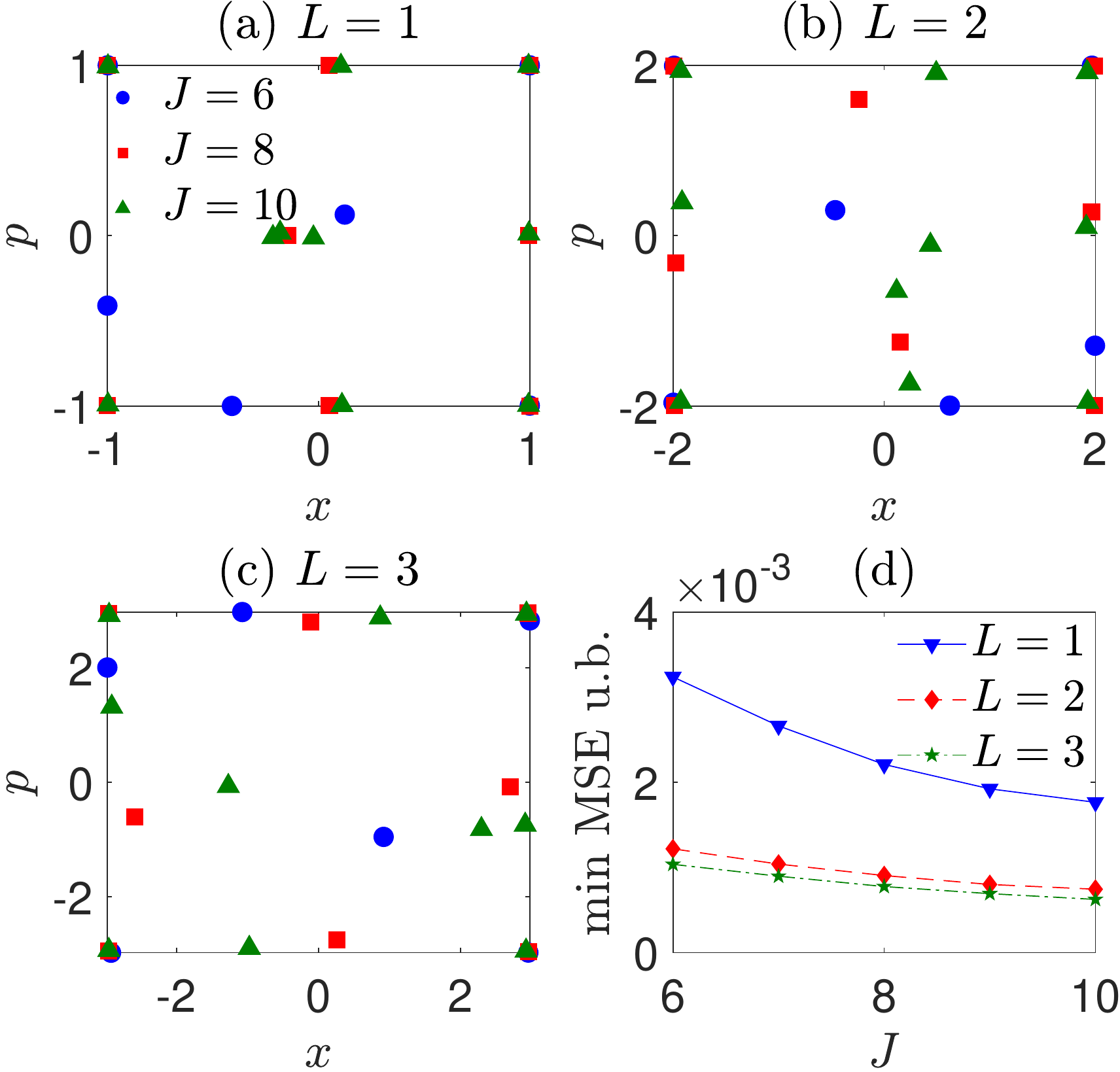}
	\caption{\label{fig:geom_states}(a,b,c)~Regarding the phase-space arrangement (represented by $x,p$ coordinate markers) of the input coherent states, the actual optimal states that collectively minimize the upper bound of \eqref{eq:asymp_cohIN_hetOUT} are geometrically positioned as far apart from each other as possible, such that some states are located in the interior of the finite-energy boundaries defined by $-L\leq x,p\leq L$. (d)~The minimum upper bound of \eqref{eq:asymp_cohIN_hetOUT} monotonically decreases with the number of input states $J$ in such geometrical sets, as expected. A saturated optimality is achieved beyond just $L=2$, beyond which increasing the laser intensity further becomes moot. For low-energy applications, $L=1$ is sufficient for precise Gaussian-process tomography.}
\end{figure}

Regardless of whether or not the TP constraint is imposed when reconstructing $\Phi$, the best performance of any Gaussian-process characterization is ultimately tied to the optimal value of the figure of merit in question. For our case, this is quantified by the minimum of the MSE. The truly optimal set of input states that minimizes the MSE according to either Eq.~\eqref{eq:asymp_cohIN_hetOUT} or \eqref{eq:asymp_cohIN_hetOUT_TP} requires the knowledge of the unknown Gaussian process. Such a set is therefore operationally unobtainable. 

We introduce a solution to approximately minimize the MSE without such knowledge by first noting that since $\dyadic{Y}\geq0$, a variant of the Cauchy--Schwarz inequality for positive matrices reads
\begin{equation}
\mathrm{MSE}\leq\dfrac{1}{N}\mathrm{Tr}\Big\{\widetilde{\dyadic{V}^-}^\dag\widetilde{\dyadic{V}^-}\Big\}\Tr{\dyadic{Y}}\approx JN\,\mathrm{Tr}\Big\{\widetilde{\dyadic{V}^-}^\dag\widetilde{\dyadic{V}^-}\Big\}\,,
\end{equation}
where the approximation in the second line of the calculation is valid for sufficiently large $M$, so that $\widetilde{p}_{jk}\ll1$ and $Y_{jk,j'k'}\approx N^2 \widetilde{p}_{jk}\delta_{j,j'}\delta_{k,k'}$. Under this approximation, it is clear that the upper bound of the asymptotic MSE is independent of $\Phi$ and is therefore a purely geometrical term that solely depends on the collective phase-space arrangement of the input coherent states. Therefore, such \emph{geometrical} sets of input states that minimize the MSE Cauchy--Schwarz upper bound can be universally defined for any Gaussian process since they are independent of the measurement data and the unknown process.

Figure~\ref{fig:geom_states} illustrates some desirable properties of geometrical input states. In particular, when $J=6$ (minimal case), the geometrical set is unique up to a collective rotation, whereas arrangements can vary for $J>6$, at times with the possibility of two or more coherent states being very close to each other owing to overcomplete redundancy. The recipe for deriving such geometrical sets of input states does not apply to the TP version of the asymptotic MSE in Eq.~\eqref{eq:asymp_cohIN_hetOUT_TP}, since the TP constraint tangles the 9 independent Gaussian parameters in a highly nonlinear way that cannot be cleanly separated from the phase-space variables ($\dyadic{V}_\textsc{tp}$ depends on these parameters).

As we shall see in Sec.~\ref{sec:results}, for heterodyne detection, such a geometrical set of coherent states on average gives a nearly-optimal MSE when the TP constraint is lifted. Moreover, these special input states can also beat optimal input states that minimizes the MSE when trace preservation is imposed for certain classes of CPTP Gaussian processes. 

\section{Performance on CPTP Gaussian processes}
\label{sec:results}

For a given CPTP Gaussian process, both its Q-function parameters $\dyadic{A}$ and $\rvec{B}$ are related to another set of parameters (refer to Appendix~\ref{app:phys_cptp}) according to the maps
\begin{align}
\dyadic{A}=&\,\lim_{t\rightarrow\infty}\,\dfrac{1}{2}\,\dyadic{U}\left[(\dyadic{1}+\TP{\dyadic{X}})\,\dyadic{\Sigma}_t\,(\dyadic{1}+\dyadic{X})+\dyadic{0}\oplus\dyadic{Y}+\dyadic{1}/2\right]^{-1}\dyadic{U}^\dag\,,\nonumber\\
\dyadic{B}=&\,2\,\dyadic{A}\,\dyadic{U}\rvec{\mu}_0\,,
\label{eq:phys_cptp}
\end{align}
where $\dyadic{U}$ is the unitary matrix defined in Sec.~\ref{sec:bkgd}. The matrices $\dyadic{X}$ and $\dyadic{Y}$ effect the general transformations $\rvec{\mu}\rightarrow\dyadic{X}\rvec{\mu}+\rvec{\mu}_0$ and $\dyadic{\Sigma}_\textsc{w}\rightarrow\dyadic{\Sigma}_\textsc{w}'=\TP{\dyadic{X}}\,\dyadic{\Sigma}_\textsc{w}\,\dyadic{X}+\dyadic{Y}$ on the mean ($\rvec{\mu}$) and covariance ($\dyadic{\Sigma}_\textsc{w}$) of the \emph{Wigner function} describing an input Gaussian state, and
\begin{equation}
\dyadic{\Sigma}_t=\dfrac{1}{2}\,\begin{pmatrix}
\cosh t & 0 & \sinh t & 0 \\
0 & \cosh t & 0 & -\sinh t \\
\sinh t & 0 & \cosh t & 0 \\
0 & -\sinh t & 0 & \cosh t
\end{pmatrix}\,.
\label{eq:SIGMA0}
\end{equation}
The matrices $\dyadic{X}$ and $\dyadic{Y}$ may then alternatively be understood as functions of a complete set of relevant operations, namely phase shift~($\phi$), displacement~($x_0,p_0$), squeezing~($r,\theta$), losses~($\chi<1$) and amplifications~($\chi>1$), and couplings to a Gaussian reservoir~($n_\textsc{t},a_\textsc{t},\theta_\textsc{t}$):
\begin{align}
\dyadic{X}=&\,\chi\dyadic{S}(r,\theta)\,\dyadic{R}(\phi)\,,\nonumber\\
\dyadic{Y} =&\, |1-\chi|^2 \dyadic{1}/2 + \frac{n_\textsc{t}}{2}\,\TP{\dyadic{R}(\theta_\textsc{t})}\begin{pmatrix}
1+a_\textsc{t} & 0\\
0 & 1-a_\textsc{t}
\end{pmatrix} \dyadic{R}(\theta_\textsc{t})   \,,\nonumber\\
\dyadic{R}(\phi)=&\,\begin{pmatrix}
\cos\phi & \sin\phi\\
-\sin\phi & \cos\phi
\end{pmatrix}\,,\nonumber\\
\dyadic{S}(r,\theta)=&\,\,\TP{\dyadic{R}(\theta)}\begin{pmatrix}
\E{r} & 0\\
0 & \E{-r}
\end{pmatrix}\dyadic{R}(\theta)\,,\nonumber\\
\rvec{\mu}_0=&\,\begin{pmatrix}
x_0\\
p_0
\end{pmatrix}\,.
\label{eq:decomposition}
\end{align}
It should be noted that although the decomposition in \eqref{eq:decomposition} preserves the symplectic character of the transformed covariance~(see Appendix~\ref{app:phys_cptp}), it is not unique. Different Gaussian operations in various orders can achieve the same physical effect, which is the reason we are reconstructing the more generally applicable parameters $\dyadic{A}$ and $\dyadic{B}$ instead of those in (\ref{eq:decomposition}), even though the latter are directly tied to actual experimental configurations.

\begin{figure}[t]
	\centering
	\includegraphics[width=0.7\columnwidth]{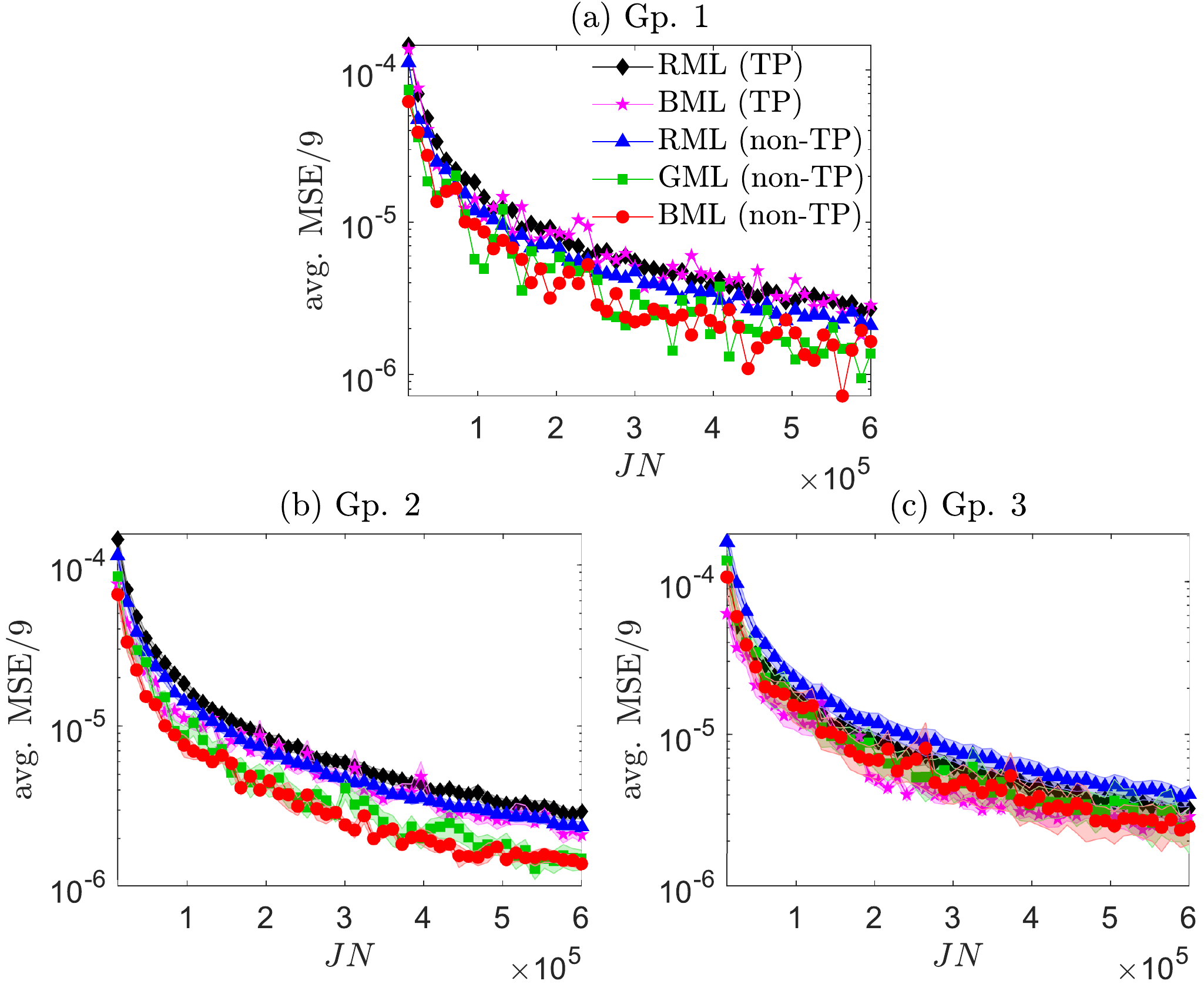}
	\caption{\label{fig:perf_TP}Performances of all five input-state strategies in reconstructing the 9 independent parameters specified in Eq.~\eqref{eq:param_TP} for various groups of unknown Gaussian CPTP processes with heterodyne detection. The MSE is properly scaled for comparison convenience with Fig.~\ref{fig:perf_nonTP}. We fixed $K=400$, $J=6$ and $L=1$ to simulate the conditions of minimal and low-energy input coherent states that are ideal for feasible tomography experiments. An average over all processes \emph{within} each group (with additional averaging over 10 random sets of input states for RML) is carried out to construct the respective $1/3$-$\sigma$ error region for the group, where the fractional $\sigma$ value is so chosen for proper illustration in the vertical logarithmic scale.}
\end{figure}

\begin{figure}[t]
	\centering
	\includegraphics[width=0.7\columnwidth]{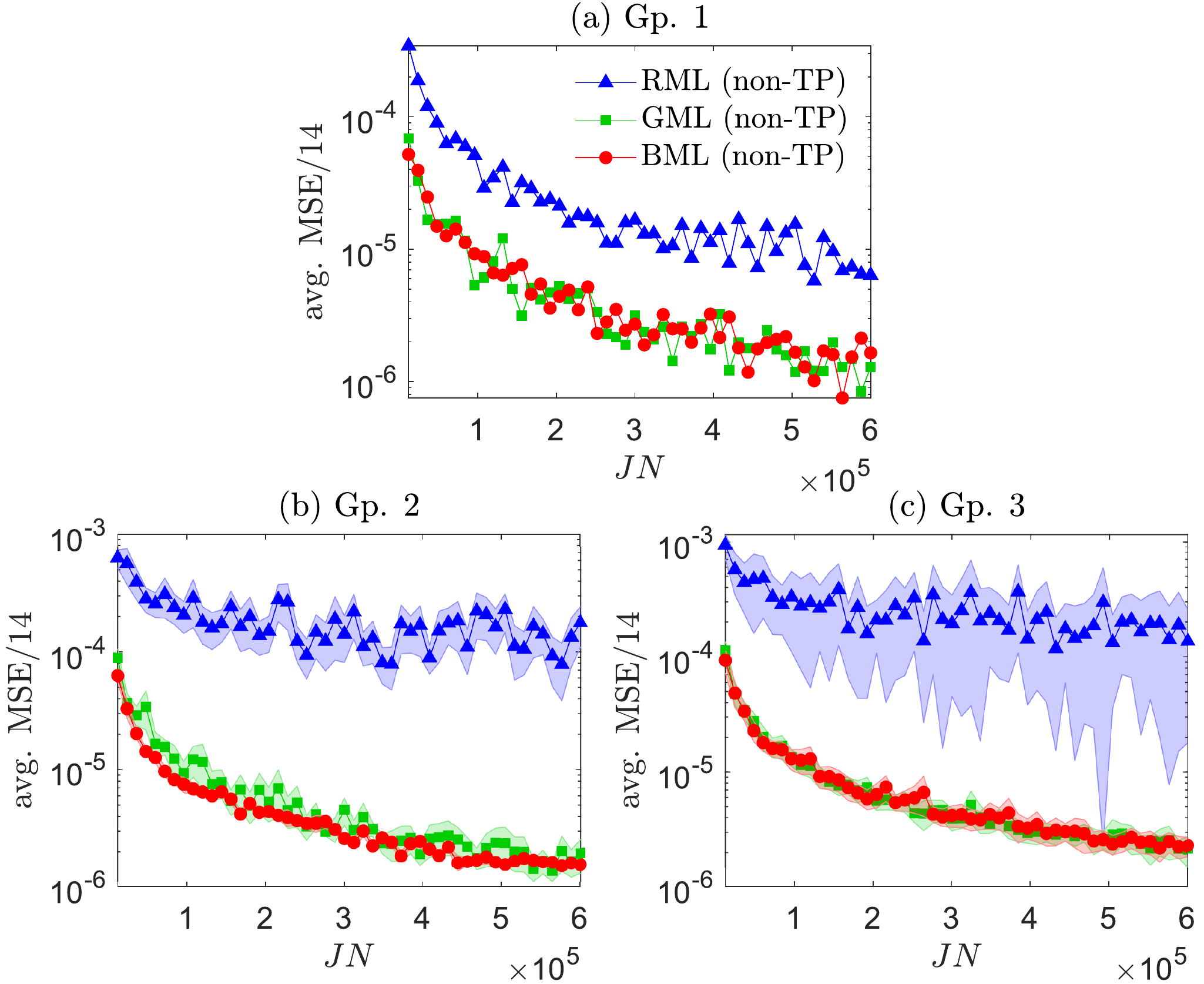}
	\caption{\label{fig:perf_nonTP}Performances of all three non-TP input-state strategies in reconstructing all 14 parameters [see Eq.~\eqref{eq:param_nonTP}] for various groups of unknown Gaussian CPTP processes with heterodyne detection. The MSE is properly scaled for comparison convenience with Fig.~\ref{fig:perf_TP} of the same values chosen for $J$, $K$ and $L$. All error regions plotted here represent 1/3-$\sigma$ standard deviation.} 
\end{figure}

\begin{figure}[t]
	\centering
	\includegraphics[width=0.7\columnwidth]{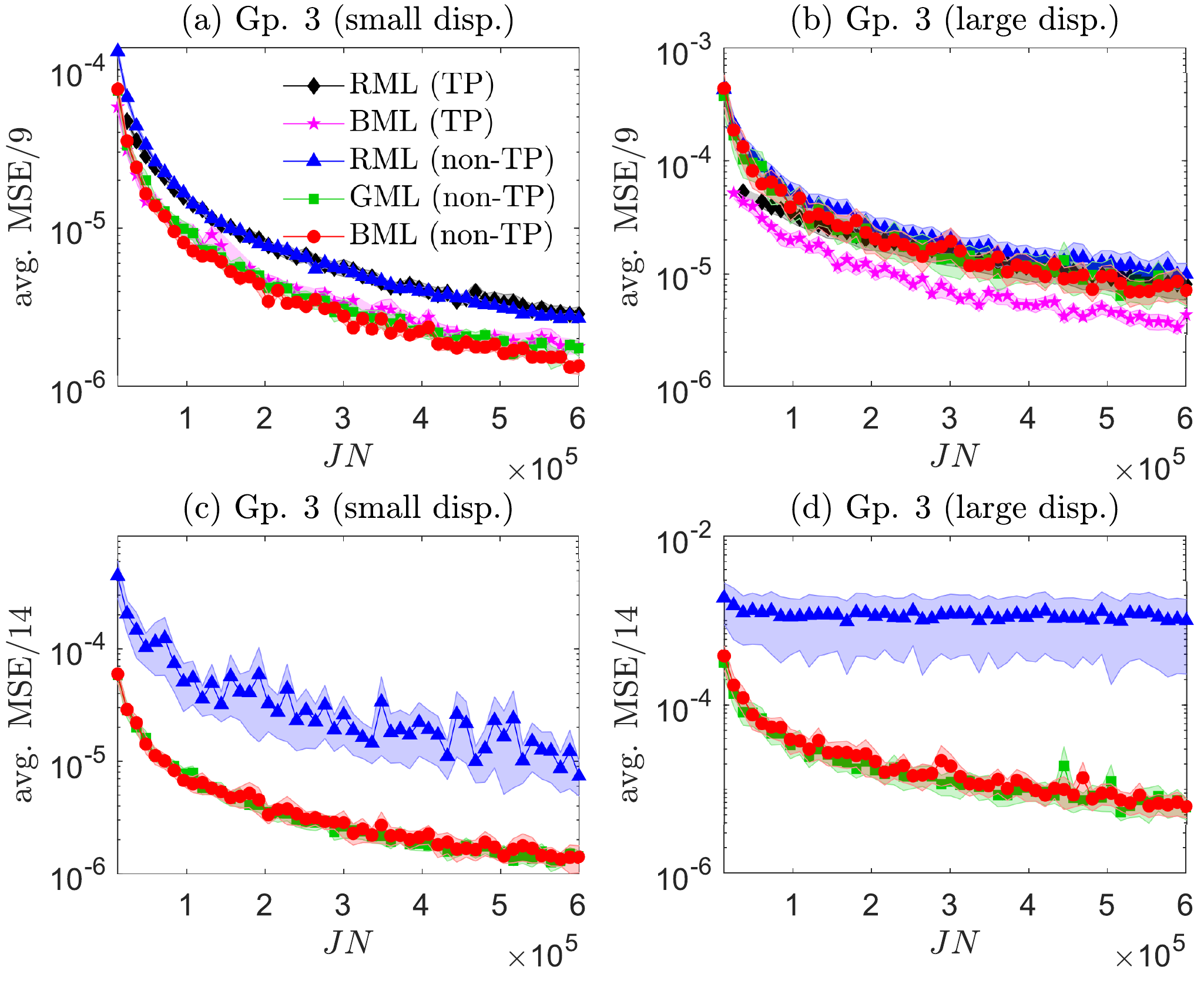}
	\caption{\label{fig:small_large}Performances of all input-state strategies for Gp.~3 random Gaussian CPTP processes with heterodyne detection. (a,c)~Small process displacements refer to those in the range $-0.5\leq x_0,p_0\leq 0.5$, and (b,d)~large displacements refer to those in $-4\leq x_0,p_0\leq 4$. All other figure specifications otherwise conform to those in Figs.~\ref{fig:perf_TP} and \ref{fig:perf_nonTP}.} 
\end{figure}

The main role of the decomposition in \eqref{eq:decomposition} is to ensure that the randomly generated processes used in our numerical experiments are those that can be found in an experimental setting. The physical parameter ranges may be fixed in the following way. The phase $\phi$ induced by the rotation operation $\dyadic{R}(\phi)$ can be completely arbitrary, $\phi\in[0,2\pi)$, and so can the phase of squeezing $\theta\in[0,\pi/2]$. The squeezing strength $r$ may be reasonably fixed to the range $r\in[0,1/3]$, which is between zero and approximately 6dB that is achievable in active research \cite{PhysRevA.90.060302,APLPhotonics.5.036104}. Displacements in optical experiments are a consequence of interaction with an external field, and may also be arbitrary. They also stand out from all the other parameters because they do not transform the covariance matrix of the input state and can be well estimated by vacuum probe states. For displacements to be comparable with the other operations carried out by the Gaussian process in strength, we shall consider the ranges $x_0,p_0\in[-2,2]$ corresponding to displacement of energy that is slightly higher than that of  squeezing. The gain of the channel $\chi$ represents both loss and amplification. Loss, caused by stray reflections, detector inefficiencies and mode mismatch, is an ever present phenomenon in quantum optical experiments, but can often be curtailed leading to transmission rates above 0.9 \cite{APLPhotonics.4.060902}. Amplification can be a result of nonlinear processes, but more often, it arises as a consequence of feed-forward with non-unit gain \cite{PhysRevLett.96.163602, IEEEquantumelectronics.9.1519}. Like displacement, it can in principle be arbitrary, but since strong amplification necessitates noise levels that destroys non-classical features of quantum states \cite{PhysRevA.90.010301}, it is usually kept low. As a conservative choice, we pick $\chi\in(0.1,1.5)$ to run all numerical experiments. The last three terms collectively characterize the added noise, which has purely detrimental effect---adding one unit of vacuum noise to both quadratures is generally sufficient to extinguish any quantum properties of the state. The asymmetry coefficient $a_\textsc{t} \in [-1,1]$ and phase $\theta_\textsc{t} \in [0,\pi/2]$ of the added noise span the practical ranges. The added-noise coupling $n_\textsc{t}\in[0,1]$ was chosen such that it describes both quantum and classical channels.

Three groups of Gaussian CPTP processes are considered here, whose physical parameters are tabulated in Tab.~\ref{tab:gps}. Monte Carlo simulations with these groups of processes are performed using heterodyne measurements, and ML reconstructions are carried out in the original matrix parametrization ($\dyadic{A},\rvec{B}$) for convenience, the specific structure of which depends on whether the TP constraint is imposed or not. Three operational input-state strategies, which are the random strategy with [RML~(TP)] and without the TP constraint [RML~(non-TP)], and the geometrical strategy (GML~(non-TP)), are evaluated by averaging the MSEs for both the 9 independent CPTP parameters and all 14 Gaussian parameters (normalized with the respective number of parameters) over all processes in each group. For benchmarking, the non-operational best strategies that respectively minimize the asymptotic MSEs in \eqref{eq:asymp_cohIN_hetOUT} [BML~(TP)] and \eqref{eq:asymp_cohIN_hetOUT_TP} [BML~(non-TP)] are also plotted. Figures~\ref{fig:perf_TP} and \ref{fig:perf_nonTP} show the performances of all five strategies. For all tested Gaussian processes within the defined physical ranges~[Figs.~\ref{fig:perf_TP}(a,b,c) and Figs.~\ref{fig:perf_nonTP}(a,b,c)], GML is the optimal choice for efficient process-parameter reconstruction with low-energy coherent states ($L=1$). 

\begin{table}[t]
	\begin{tabular}{rlcccccccccl}
		&no.& {$\phi$} & $r$ & $\theta$ & $x_0$ & $p_0$ & $\chi$ & $n_\textsc{t}$ & $a_\textsc{t}$ & $\theta_\textsc{t}$&\\
		\cline{2-11}\\[-5ex]
		\cline{2-11}\\[-3ex]
		\begin{rotate}{90}\!\!\!\!{\bf Gp.~1}\end{rotate}\quad\,&$\bm{1}$& $0$ & $0$ & $0$ & $0$ & $0$ & $1$ & $0$ & $0$ & $0$&(idle)\\[1ex]
		\cline{2-11}\\[-3ex]
		& $\bm{1}$ &  $*$ & $0$ & $0$ & $0$ & $0$ & $1$ & $0$ & $0$ & $0$&(phase shifter)\\
		&$\bm{2}$ & $0$ & $*$ & $*$ & $0$ & $0$ & $1$ & $0$ & $0$ & $0$&(squeezer)\\
		&$\bm{3}$ & $0$ & $0$ & $0$ & $*$ & $*$ & $1$ & $0$ & $0$ & $0$&(displacer)\\
		\begin{rotate}{90}\!\!\!\!{\bf Gp.~2}\end{rotate}\quad\,&$\bm{4}$ & $0$ & $0$ & $0$ & $0$ & $0$ & $*$ & $0$ & $0$ & $0$&(gain)\\
		&$\bm{5}$ & $0$ & $0$ & $0$ & $0$ & $0$ & $1$ & $*$ & $0$ & $0$&(symmetric noise)\\
		&$\bm{6}$ & $0$ & $0$ & $0$ & $0$ & $0$ & $1$ & $*$ & $*$ & $*$&(asymmetric noise)\\[1ex]
		\cline{2-11}\\[-3ex]
		\begin{rotate}{90}\!\!\!\!{\bf Gp.~3}\end{rotate}\quad\,&$\bm{1}$-$\bm{10}$& $*$ & $*$ & $*$ & $*$ & $*$ & $*$ & $*$ & $*$ & $*$&(arbitrary)\\[1ex]
		\cline{2-11}\\[-5ex]
		\cline{2-11}
	\end{tabular}
	\caption{\label{tab:gps}Physical parameters characterizing the five different groups of CPTP processes invoked in the simulations. The wildcard~$*$ denotes a randomly generated value within the corresponding interval for the parameter, as stated in Sec.~\ref{sec:results}. Gp.~1 consists of the singular idle process ($\Phi[\rho]=\rho$), Gp.~2 contains 6 random processes that each represents one basic type of Gaussian operation or noise character (examples 5 and 6 respective coincide with additive symmetric and asymmetric noise arising from a thermal bath). Finally, Gp.~3 consists of 10 completely arbitrary processes.}
\end{table}

Interesting dynamics reveal themselves for the completely random CPTP processes in Gp.~3, where we find that, again, GML beats all strategies when all parameters are arbitrarily chosen with the displacement ranges $-2\leq x_0,p_0\leq2$ obeyed. For the tested random processes, Figs.~\ref{fig:perf_TP}(c) and \ref{fig:perf_nonTP}(c) highlight rather comparable performances between GML and the optimal BML~(TP). However, as shown in Fig.~\ref{fig:small_large}, if we enlarge the displacement ranges, we find that the best TP input-state strategy can outperform the rest. On the other hand, when these ranges are reduced, GML reconstructs process parameters with much better accuracies than for default ranges as in Figs.~\ref{fig:perf_TP}(c) and \ref{fig:perf_nonTP}(c). This leads us to conjecture that the additional reconstruction bias introduced by the TP constraint, on top of that from the CP constraint, apparently has beneficial merits for more correctly singling out estimators that are near the true CPTP Gaussian processes that perform large displacing operations; whereas processes with stronger second-moment manipulating features relative to first-moment displacements are still better characterized with non-TP input-state strategies as they appear to be more robust against noise, in which case GML is the optimal choice.

\section{Conclusion}

There have been many studies related to achieving the quantum limits of parameter estimation. We have taken a different route instead and investigated an experimentally feasible and tomographically efficient way to characterize Gaussian quantum processes. Using heterodyne measurements and input coherent states, we introduced a simple strategy of constructing geometrical sets of input coherent states that effectively optimizes the process parameters' mean squared-error. These geometrical input states are demonstrated to outperform the best nonadaptive input-state strategy in terms of the mean squared-error for typical CPTP processes that do not carry out large displacement operations. This permits us to utilize coherent states of low energies as sufficient convenient resources to achieve very low mean squared-errors in the reconstructed process parameters. 

We also observe that if the unknown Gaussian process has large displacing features on input states, input-state strategies that imposes the trace-preserving constraint, on average, give more accurate process estimators than the geometrical strategy where this constraint is relaxed. In the course of acquiring these results, we also obtained asymptotic analytical expressions for the process-parameter mean squared error that were previously not discussed in the quantum process tomography literature to the authors' knowledge.

The next natural step would be to investigate the extent of enhancement when the input coherent states are squeezed. Preliminary studies show that the asymptotic mean squared-error expressions with heterodyning apparently becomes exceedingly complicated, such that there is currently no straightforward recipe to construct the geometrical sets of input states discussed here. Whether there exist output-state measurements that are more compatible with squeezed input states in probing Gaussian processes other than heterodyning is an interesting open question.

\acknowledgments{Y.S.T., S.S. and H.J. acknowledge support by the National Research Foundation of Korea (NRF) (Grant Nos. NRF-2019R1A6A1A10073437, NRF-2018K2A9A1A06069933, NRF-2019M3E4A1080074 and NRF-2020R1A2C1008609); K.P. and P.M. were supported by project 19-19722J of the Grant Agency of Czech Republic (GA\v{C}R).}

\appendix

\section{Trace-preserving Gaussian processes}
\label{app:TP}

The TP constraint imposed on a quantum process $\Phi$ is defined by the partial-trace relation $\mathrm{tr}_2\{\rho_\Phi\}=1$ for its Choi--Jamio{\l}kowski operator $\rho_\Phi$. Using the overcompleteness property of the coherent states, $\int(\D \beta)\ket{\beta}\bra{\beta}/\pi=1$, this partial-trace relation is expressed as
\begin{equation}
\int\dfrac{(\D\beta')}{\pi}\,\bm{:}\exp\!\left(-(\rvec{a}^\dag\,\,\rvec{\beta}'^\dag)\,\dyadic{A}\begin{pmatrix}
\rvec{a}\\
\rvec{\beta}'
\end{pmatrix}+\rvec{B}^\dag\begin{pmatrix}
\rvec{a}\\
\rvec{\beta}'
\end{pmatrix}+c_0\right)\!\bm{:}\,\,=1\,,
\label{eq:pt_normord}
\end{equation}
where $\rvec{a}=\TP{(a\,\,\,a^\dag)}$ is the column of ladder operators. The normal-order expression in \eqref{eq:pt_normord} may be unraveled with two other overcomplete sets of coherent states, 
\begin{align}
&\,\int\dfrac{(\D\alpha)}{\pi}\int\dfrac{(\D\alpha')}{\pi}\ket{\alpha}\E{-\frac{1}{2}|\alpha|^2-\frac{1}{2}|\alpha'|^2+\alpha^*\alpha'}\bra{\alpha'}\nonumber\\
&\,\times\int\dfrac{(\D\beta')}{\pi}\,\exp\!\left(-(\widetilde{\widetilde{\rvec{\alpha}}}\,\,\,\rvec{\beta}'^\dag)\,\dyadic{A}\begin{pmatrix}
\widetilde{\rvec{\alpha}}\\
\rvec{\beta}'
\end{pmatrix}+\rvec{B}^\dag\begin{pmatrix}
\widetilde{\rvec{\alpha}}\\
\rvec{\beta}'
\end{pmatrix}+c_0\right)=1\,,\nonumber\\
&\,\widetilde{\rvec{\alpha}}=\TP{(\alpha'\,\,\,\alpha^*)}\,,\nonumber\\
&\,\widetilde{\widetilde{\rvec{\alpha}}}=\TP{\widetilde{\rvec{\alpha}}}\dyadic{\sigma}_x\,,\,\,\,\dyadic{\sigma}_x\,\,\widehat{=}\,\begin{pmatrix}
0&1\\1&0
\end{pmatrix}\,.
\end{align}

The Gaussian integration in $\beta'$ can be evaluated by using the general formula
\begin{align}
&\,\int\dfrac{(\D\beta')}{\pi}\,\E{-a|\beta'|^2+b_1\beta'+b_2\beta'^*+c_1\beta'^2+c_2{\beta'^*}^2}\nonumber\\
=&\,\dfrac{1}{\sqrt{a^2-4c_1c_2}}\exp\left(\dfrac{ab_1b_2+c_1b_2^2+c_2b_1^2}{a^2-4c_1c_2}\right)
\label{eq:gauss_complex}
\end{align}
for $\RE{a}\geq|c_1+c^*_2|$ (see \cite{Teo:2015qs,Perina:2001aa}). At times, it is much easier to cope with the matrix form of the integration:
\begin{align}
&\,\int\dfrac{(\D\beta')}{\pi}\,\E{-\rvec{\beta'}^\dag\dyadic{M}\rvec{\beta'}+\TP{\rvec{v}}\rvec{\beta}'}=\dfrac{1}{2\sqrt{\DET{\dyadic{M}}}}\E{\TP{\rvec{v}}\dyadic{M}^{-1}\dyadic{\sigma}_x\rvec{v}/4}\,.
\label{eq:gauss_complex2}
\end{align}
After an application of \eqref{eq:gauss_complex2} and some reorganization of terms, we have
\begin{align}
1=&\,\int\dfrac{(\D\alpha)}{\pi}\int\dfrac{(\D\alpha')}{\pi}\ket{\alpha}\inner{\alpha}{\alpha'}\,\E{-\widetilde{\widetilde{\rvec{\alpha}}}\,\dyadic{W}\widetilde{\rvec{\alpha}}+\rvec{y}^\dag\widetilde{\rvec{\alpha}}+w_0}\bra{\alpha'}\,,\nonumber\\
\dyadic{W}=&\,\dyadic{A}_1-\dyadic{A}_2\,\dyadic{A}_3^{-1}\,\dyadic{A}_2^\dag\,,\nonumber\\
\rvec{y}=&\,\rvec{b}_1-\dyadic{A}_2\,\dyadic{A}_3^{-1}\rvec{b}_2\,,\nonumber\\
w_0=&\,c_0-\log(2\sqrt{\DET{\dyadic{A}_3}})+\dfrac{1}{4}\rvec{b}^\dag_2\,\dyadic{A}_3^{-1}\rvec{b}_2\,.
\label{eq:Gauss_res}
\end{align}

It therefore follows that the sufficient conditions for $\Phi$ to be TP are $\dyadic{W}=\dyadic{0}$, $\rvec{y}=\rvec{0}$ and $w_0=0$. To show that these are necessary conditions, we rewrite \eqref{eq:Gauss_res} as
\begin{equation}
\bm{:}\E{-\rvec{a}^\dag\dyadic{W}\rvec{a}+\rvec{y}^\dag\rvec{a}+w_0}\bm{:}\,\,=1
\label{eq:normord}
\end{equation}
and make use of Eqs.~(1.11)--(1.13) in~\cite{Agrawal:1977aa} to convert the normal-order form on the left-hand side of \eqref{eq:normord} into
\begin{equation}
\bm{:}\E{-\rvec{a}^\dag\dyadic{W}\rvec{a}+\rvec{y}^\dag\rvec{a}}\bm{:}\,\,=\dfrac{1}{\kappa}\,\E{-\rvec{a}^\dag\dyadic{W}\dyadic{R}_1\dyadic{T}^{-1}\dyadic{R}_1\rvec{a}+\rvec{y}^\dag\dyadic{R}_1\dyadic{T}^{-1}\dyadic{R}_1\rvec{a}}\,,\label{eq:Gauss_normord}
\end{equation}
where
\begin{align}
\kappa=&\,\DET{\mathrm{sinc}(\dyadic{\sigma}_y\,\dyadic{\xi})}^{-1/2}\DET{\dyadic{T}}^{1/2}\exp(\TP{\rvec{\eta}}(\dyadic{T-\dyadic{1}})\dyadic{\xi}^{-1}\rvec{\eta})\,,\nonumber\\
\dyadic{T}=&\,\left[\cos(\dyadic{\sigma}_y\,\dyadic{\xi})-\I\dyadic{\sigma}_z\sin(\dyadic{\sigma}_y\,\dyadic{\xi})\right][\mathrm{sinc}(\dyadic{\sigma}_y\,\dyadic{\xi})]^{-1}\,,\nonumber\\
\dyadic{\xi}=&\,-\TP{\dyadic{R}}_2\,\dyadic{W}\,\dyadic{R}_1\,,\quad\TP{\rvec{\eta}}=\rvec{y}^\dag\dyadic{R}_1\,,\nonumber\\
\dyadic{R}_1=&\,\begin{pmatrix}
1&0&0&0\\
0&0&1&0\\
0&1&0&0\\
0&0&0&1
\end{pmatrix}\,,\quad\dyadic{R}_2=\begin{pmatrix}
0&0&1&0\\
1&0&0&0\\
0&0&0&1\\
0&1&0&0
\end{pmatrix}\,.
\end{align}
It is now obvious that the earlier sufficient conditions are also necessary for the Gaussian form on the right-hand side of \eqref{eq:Gauss_normord} to be the identity operator, as $\dyadic{R}_1\dyadic{T}_1^{-1}\dyadic{R}_1$ is non-singular. 

\section{Asymptotic statistics of logarithmic data}
\label{app:asymp_log}

From the inversion solution $\widehat{\rvec{x}}=\dyadic{V}^{-}\rvec{u}$, identifying the large-data characteristics concerning $-\log\nu_{jk}$ is crucial in deriving the asymptotic behavior of the parameter MSE stated in \eqref{eq:asymp_cohIN_hetOUT}. We first realize that in the absence of systematic errors and under low fluctuation in $\gamma_j$, $\nu_{jk}\equiv\gamma_j\widetilde{\nu}_{jk}\rightarrow\gamma_j\widetilde{p}_{jk}$ in the regime $N\gg1$, where $\sum_k\widetilde{\nu}_{jk}=1$ constitute the actual (normalized) data collected after performing heterodyning on the $j$th output state. Upon denoting $\Delta_{jk}=-\log \nu_{jk}+\log p_{jk}$, this invites us to expand $\Delta_{jk}\Delta_{j'k'}$ that appears in the MSE $(\widehat{\rvec{x}}-\rvec{x})^2$ of interest about $\widetilde{\nu}_{jk}-\widetilde{p}_{jk}$ to facilitate the statistical averaging. 

We begin with the basic Taylor expansion
\begin{equation}
\Delta_{jk}\approx-\dfrac{1}{\widetilde{p}_{jk}}(\widetilde{\nu}_{jk}-\widetilde{p}_{jk})+\dfrac{1}{2\widetilde{p}^2_{jk}}(\widetilde{\nu}_{jk}-\widetilde{p}_{jk})^2\,.
\end{equation}
After binning the sampled data over the continuous phase space into an $M\times M$ ($x,p$) grid, the $j$th output Q~function may be analyzed in terms of an $(K=M^2)$-outcome multinomial distribution defined by the normalized $\{\widetilde{p}_{jk}\}$ to arbitrarily high accuracy if $M\gg1$. Under multinomial statistics, we have the well-known formula
\begin{equation}
\overline{(\widetilde{\nu}_{jk}-\widetilde{p}_{jk})(\widetilde{\nu}_{j'k'}-\widetilde{p}_{j'k'})}=\dfrac{\delta_{j,j'}}{N}(\delta_{k,k}\widetilde{p}_{jk}-\widetilde{p}_{jk}\widetilde{p}_{j'k'})
\end{equation}
for the covariance in view of the fact that data arising from different output states are statistically independent.

The last detail that we need to pay attention to is the particular step in the inversion procedure for obtaining the estimator $\widehat{\rvec{x}}$ where outcomes with zero counts are discarded in order for all the logarithms of $\widehat{\rvec{u}}$ to be well-defined. This implies that there is nonzero contributions to $\overline{\Delta_{jk}\Delta_{j'k'}}$ only when $\widetilde{\nu}_{jk}>0$. The probability for this occurring is $1-(1-\widetilde{p}_{jk})^N$, which is found by noting that the probability $p_{0,jk}=\mathrm{prob}(n_{jk}=0|\{\widetilde{p}_{jk}\})$ that the frequency of the $k$th outcome is zero in a multinomial distribution is
\begin{equation}
p_{0,jk}=\sum_{\{n_{l\neq k}\}}\dfrac{N!}{\prod_{l'\neq k}n_{jl'}!}\,\prod_{l''\neq k}\widetilde{p}^{n_{jl''}}_{jl''}=(1-\widetilde{p}_{jk})^N\,.
\end{equation}
With this,
\begin{equation}
\overline{\Delta_{jk}\Delta_{j'k'}}\approx\dfrac{(1-p_{0,jk})(1-p_{0,j'k'})}{N}\,\delta_{j,j'}\left(\dfrac{\delta_{k,k}}{\widetilde{p}_{jk}}-1\right)\,,
\end{equation}
where regularity is guaranteed, as it should, because of the limit
\begin{equation}
\dfrac{[1-(1-\widetilde{p}_{jk})^N]^2}{\widetilde{p}_{jk}}\approx N^2 \widetilde{p}_{jk}\rightarrow 0
\end{equation}
as $\widetilde{p}_{jk}\rightarrow 0$. The final expressions stated in \eqref{eq:asymp_cohIN_hetOUT} hence follow suit.

The derivation of the MSE expression for the ML reconstruction scheme is only slightly more technical. In calculating the increment $\updelta\log\mathcal{L}$, variations in the respective parameter matrices and column lead to
\begin{align}
\updelta\log p_{jk}=&\,\,\Tr{\dyadic{M}^\dag_{1,jk}\updelta\dyadic{A}_3}+\Tr{(\dyadic{M}_{2,jk}+\dyadic{M}_{2,jk}^\dag)\updelta\dyadic{A}_2}\nonumber\\
&\,+\rvec{M}_{3,jk}^\dag\updelta\rvec{b}_2\,,\nonumber\\
\dyadic{M}_{1,jk}=&\,-\rvec{z}_k\,\rvec{z}_k^\dag+\dyadic{A}_3^{-1}\dyadic{A}_2^\dag\,\rvec{\alpha}_j\rvec{\alpha}_j^\dag\,\dyadic{A}_2\,\dyadic{A}_3^{-1}+\frac{1}{2}\,\dyadic{A}_3^{-1}\nonumber\\
&\,-\dyadic{A}_3^{-1}\rvec{b}_2\,\rvec{\alpha}_j^\dag\,\dyadic{A}_2\,\dyadic{A}_3^{-1}+\frac{1}{4}\,\dyadic{A}_3^{-1}\rvec{b}_2\,\rvec{b}_2^\dag\,\dyadic{A}_3^{-1}\,,\nonumber\\
\dyadic{M}_{2,jk}=&\,-\rvec{\alpha}_j\,\rvec{\alpha}_j^\dag\,\dyadic{A}_2^\dag\,\dyadic{A}_3^{-1}+\frac{1}{2}\,\dyadic{A}_3^{-1}\rvec{b}_2\,\rvec{\alpha}_j^\dag+\dyadic{\sigma}_x\,\rvec{\alpha}_j\,\rvec{z}_k^\dag\,,\nonumber\\
\rvec{M}_{3,jk}=&\,\,\rvec{z}_k+\,\dyadic{A}_3^{-1}\dyadic{A}_2\,\rvec{\alpha}_j-\frac{1}{2}\,\dyadic{A}_3^{-1}\rvec{b}_2\,,
\end{align}
where the gradient components $\dyadic{M}_{1,jk}$, $\dyadic{M}_{2,jk}$ and $\rvec{M}_{3,jk}$ can be derived using the simple variational identities $\updelta\dyadic{Y}^{-1}=-\dyadic{Y}^{-1}\updelta\dyadic{Y}\,\dyadic{Y}^{-1}$ and $\updelta\DET{\dyadic{Y}}=\DET{\dyadic{Y}}\,\Tr{\dyadic{Y}^{-1}\,\updelta\dyadic{Y}}$ for any invertible $\dyadic{Y}$. At this stage, we introduce the vectorization notation $\sket{\dyadic{Y}}$ that refers to the column formed by stacking all columns of $\dyadic{Y}$ in the computational-basis representation, and its dual $\sket{\dyadic{Y}}^\dag\equiv\sbra{\dyadic{Y}^\dag}$. These higher-dimensional objects relate to the trace inner product of two matrices $\dyadic{Y}_1$ and $\dyadic{Y}_2$ \emph{via} $\Tr{\dyadic{Y}_1^\dag\,\dyadic{Y}_2}=\sinner{\dyadic{Y}_1^\dag}{\dyadic{Y}_2}$. Next, we identify the essential transformations to calculate the $9\times1$ $\updelta\rvec{x}$:
\begin{align}
\sket{\updelta\dyadic{A}_3}=&\,\,\dyadic{E}_1\,\TP{(\updelta a_2\,\,\,\updelta(c_2)_\mathrm{r}\,\,\,\updelta(c_2)_\mathrm{i})}\,,\nonumber\\
\sket{\updelta\dyadic{A}_2}=&\,\,\dyadic{E}_2\,\TP{(\updelta (g_1)_\mathrm{r}\,\,\,\updelta (g_1)_\mathrm{i}\,\,\,\updelta(g_2)_\mathrm{r}\,\,\,\updelta(g_2)_\mathrm{i})}\,,\nonumber\\
\updelta\rvec{b}_2=&\,\,\dyadic{E}_3\,\TP{(\updelta(b_2)_\mathrm{r}\,\,\,\updelta(b_2)_\mathrm{i})}\,,\nonumber\\
\dyadic{E}_1=&\,\begin{pmatrix}
\frac{1}{2} & 0 & 0\\
0 & -1 & -\I\\
0 & -1 & \I\\
\frac{1}{2} & 0 & 0
\end{pmatrix}\,,\,\,\,\,\,\dyadic{E}_2=\,\frac{1}{2}\begin{pmatrix}
0 & 0 & 1 & \I\\
1 & \I & 0 & 0\\
1 & -\I & 0 & 0\\
0 & 0 & 1 & -\I
\end{pmatrix}\,,\nonumber\\
\dyadic{E}_3=&\,\begin{pmatrix}
1 & \I\\
1 & -\I
\end{pmatrix}\,.
\end{align}
All these auxiliary matrices form the components needed to establish Eq.~\eqref{eq:asymp_cohIN_hetOUT_TP}.

\section{Physical features of CPTP Gaussian processes}
\label{app:phys_cptp}

In phase-space representation, the second-moment matrix $\dyadic{A}'$ of the process Q~function $\exp\,(-\TP{\rvec{R}} \dyadic{A}'\, \rvec{R}+\TP{\rvec{B}'}\rvec{R}+c_0)$ is related to the Q-function covariance $\dyadic{\Sigma}_\textsc{q}$ as $\dyadic{A}'=\dyadic{\Sigma}^{-1}_\textsc{q}/2$ by definition. Since the Q~function is a Gaussian convolution of the Wigner function, this Q-function covariance can in turn be obtained from the Wigner-function covariance~$\dyadic{\Sigma}_\textsc{w}$ in accordance with $\dyadic{\Sigma}_\textsc{q}=\dyadic{\Sigma}_\textsc{w}+\dyadic{1}/2$. We may equivalently regard $\dyadic{\Sigma}_\textsc{w}$ as the output of a general covariance transformation on the otherwise idle channel covariance $\dyadic{\Sigma}_0$:
\begin{equation}
\dyadic{\Sigma}_\textsc{w}=(\dyadic{1}\oplus\,\TP{\dyadic{X}})\,\dyadic{\Sigma}_0\,(\dyadic{1}\oplus\,\dyadic{X})+\dyadic{0}\oplus\dyadic{Y}\,,
\label{eq:02wig}
\end{equation} 
where $\dyadic{X}$ and $\dyadic{Y}$ precisely effect such a transformation on the covariance matrix $\dyadic{\Sigma}$ defined for the Gaussian Wigner function of the input state inasmuch as $\dyadic{\Sigma}_\textsc{w}'=\TP{\dyadic{X}}\dyadic{\Sigma}_\textsc{w}\dyadic{X}+\dyadic{Y}$. The matrices $\dyadic{X}$ and $\dyadic{Y}$ are functions of the operational parameters defined in Sec.~\ref{sec:results}. Although their functional forms may not be uniquely established, the transformed covariance $\dyadic{\Sigma}_\textsc{w}'$ must satisfy the inequality
\begin{equation}
\dyadic{\Sigma}_\textsc{w}'+\I\,\dyadic{\Omega}/2\geq\dyadic{0}\quad\mathrm{for}\quad\dyadic{\Omega}=\begin{pmatrix}
0 & 1\\
-1 & 0
\end{pmatrix}\,,
\label{eq:XYsymp}
\end{equation} 
which preserves Heisenberg's uncertainty relations~\cite{Weedbrook:2012ag}. To confirm that the decomposition in \eqref{eq:decomposition} for $\dyadic{X}$ and $\dyadic{Y}$ precisely obeys this inequality, recall that $\dyadic{X}=\chi\,\dyadic{S}\,\dyadic{R}$ is proportionally symplectic---$\TP{\dyadic{X}}\dyadic{\Omega}\,\dyadic{X}=\chi^2\TP{\dyadic{R}}\TP{\dyadic{S}}\dyadic{\Omega}\,\dyadic{S}\,\dyadic{R}=\chi^2\,\dyadic{\Omega}$. On the other hand, $\dyadic{Y}=|1-\chi^2|\dyadic{1}/2+\dyadic{Y}_0$ is a linear combination of a symplectic multiple and a positive matrix $\dyadic{Y}_0$. Thus,
\begin{align}
&\,\dyadic{\Sigma}_\textsc{w}'+\I\,\dyadic{\Omega}/2\nonumber\\
=&\,|1-\chi^2|\dyadic{1}/2-\I(\chi^2-1)\,\dyadic{\Omega}/2+\TP{\dyadic{X}}(\dyadic{\Sigma}+\I\,\dyadic{\Omega}/2)\,\dyadic{X}+\dyadic{Y}_0\,,
\end{align}
where the final two matrices are clearly positive. The first two matrices combine to give a projector multiple for any $\chi$. In terms of the standard Pauli matrix $\dyadic{\sigma}_y$, they are $(1-\chi^2)(1-\dyadic{\sigma}_y)/2$ and $(\chi^2-1)(1+\dyadic{\sigma}_y)/2$ for the respective ranges $\chi<1$ and $\chi>1$.  

Next, to derive the relation between the pairs $(\dyadic{A},\rvec{B})$ and $(\dyadic{X},\dyadic{Y})$, we note that $\dyadic{A}'=\dyadic{A}'_0$ for the idle process ($\dyadic{B}'=\dyadic{0}$) is given by
\begin{equation}
\dyadic{A}'_0=\dfrac{1}{2}\begin{pmatrix}
1 & 0 & -1 & 0\\
0 & 1 & 0 & 1\\
-1 & 0 & 1 & 0\\
0 & 1 & 0 & 1
\end{pmatrix}\,,
\end{equation}
which is rank-2 due to the CPTP character~(recall Sec.~\ref{subsec:ML_TP}). Moreover, $\dyadic{A}'_0$ is a rank-2 projector---$\dyadic{A}'^2_0=\dyadic{A}'_0$. 

There is therefore strictly no valid $\dyadic{\Sigma}_0$ for the idle process as $\dyadic{A}'_0$ is not invertible in the usual sense. The validity of \eqref{eq:02wig} would demand an understanding that $\dyadic{A}$ is truly invertible as soon as the TP constraint is violated, and that all parameter variations in $\dyadic{A}$ should encounter no discontinuities when transiting between non-TP and TP subspaces. Upon recognizing that the Choi-Jamio{\l}kowski operator for the idle channel is represented by the superposition $\sum^\infty_{n=0}\ket{nn}$, we may consider a more physical realization using a two-mode squeezed vacuum state of real finite squeezing strength $\omega\gg1$:
\begin{equation}
\Phi^{\mathrm{idle}}\,\,\,\widehat{=}\sum^\infty_{n,n'=0}\ket{nn}\bra{n'n'}\rightarrow\Phi^{\mathrm{idle}}_\omega\,\,\,\widehat{=}\sum^\infty_{n,n'=0}\ket{nn}(\tanh \omega)^{n+n'}\bra{n'n'}\,.
\end{equation}
It follows that the relevant $\omega$-deformed Q~function spells 
\begin{equation}
Q_\Phi(\omega)=\opinner{\alpha,z}{\Phi^{\mathrm{idle}}_\omega}{\alpha,z}=\E{-|\alpha|^2-|z|^2+(\alpha z+\alpha^* z^*)\tanh \omega}\,,
\end{equation}
from whence one reads off the full-rank matrices
\begin{align}
\dyadic{A}=&\,\dfrac{1}{2}\begin{pmatrix}
1 & 0 & 0 & -\tanh w\\
0 & 1 & -\tanh \omega & 0\\
0 & -\tanh \omega & 1 & 0\\
-\tanh \omega & 0 & 0 & 1
\end{pmatrix}\,,\nonumber\\
\qquad\dyadic{A}'=&\,\dfrac{1}{2}\begin{pmatrix}
1 & 0 & -\tanh \omega & 0\\
0 & 1 & 0 & \tanh \omega\\
-\tanh \omega & 0 & 1 & 0\\
0 & \tanh \omega & 0 & 1
\end{pmatrix}\,.
\end{align}
Obviously, in the limit $\omega\rightarrow\infty$, we revert to $\dyadic{A}'\rightarrow\dyadic{A}'_0$. Yet, any finite $\omega$ leads to a well-defined $\dyadic{\Sigma}_t=(2\,\dyadic{A}')^{-1}|_{t=2\omega}-\dyadic{1}/2$ in Eq.~\eqref{eq:SIGMA0} after the substitution $t=2\omega$. That the inverse of $\dyadic{A}'_0$ for the idle process is ill-defined should now be even clearer from $\lim_{t\rightarrow\infty}\dyadic{\Sigma}_t=\dyadic{\Sigma}_0$. Despite this divergence, the limit expression in \eqref{eq:phys_cptp} must tend to a finite matrix $\dyadic{A}$, as to be expected from the physical point of view. 

\input{gauss_proc_tomo.bbl}

\end{document}

%% file: gauss_proc_tomo.bbl
%

%% file: gauss_proc_tomo.bbl
\begin{thebibliography}{55}%
\makeatletter
\providecommand \@ifxundefined [1]{%
 \@ifx{#1\undefined}
}%
\providecommand \@ifnum [1]{%
 \ifnum #1\expandafter \@firstoftwo
 \else \expandafter \@secondoftwo
 \fi
}%
\providecommand \@ifx [1]{%
 \ifx #1\expandafter \@firstoftwo
 \else \expandafter \@secondoftwo
 \fi
}%
\providecommand \natexlab [1]{#1}%
\providecommand \enquote  [1]{``#1''}%
\providecommand \bibnamefont  [1]{#1}%
\providecommand \bibfnamefont [1]{#1}%
\providecommand \citenamefont [1]{#1}%
\providecommand \href@noop [0]{\@secondoftwo}%
\providecommand \href [0]{\begingroup \@sanitize@url \@href}%
\providecommand \@href[1]{\@@startlink{#1}\@@href}%
\providecommand \@@href[1]{\endgroup#1\@@endlink}%
\providecommand \@sanitize@url [0]{\catcode `\\12\catcode `\$12\catcode
  `\&12\catcode `\#12\catcode `\^12\catcode `\_12\catcode `\%12\relax}%
\providecommand \@@startlink[1]{}%
\providecommand \@@endlink[0]{}%
\providecommand \url  [0]{\begingroup\@sanitize@url \@url }%
\providecommand \@url [1]{\endgroup\@href {#1}{\urlprefix }}%
\providecommand \urlprefix  [0]{URL }%
\providecommand \Eprint [0]{\href }%
\providecommand \doibase [0]{http://dx.doi.org/}%
\providecommand \selectlanguage [0]{\@gobble}%
\providecommand \bibinfo  [0]{\@secondoftwo}%
\providecommand \bibfield  [0]{\@secondoftwo}%
\providecommand \translation [1]{[#1]}%
\providecommand \BibitemOpen [0]{}%
\providecommand \bibitemStop [0]{}%
\providecommand \bibitemNoStop [0]{.\EOS\space}%
\providecommand \EOS [0]{\spacefactor3000\relax}%
\providecommand \BibitemShut  [1]{\csname bibitem#1\endcsname}%
\let\auto@bib@innerbib\@empty
\bibitem [{\citenamefont {Braunstein}\ and\ \citenamefont {van
  Loock}(2005)}]{Braunstein:2005aa}%
  \BibitemOpen
  \bibfield  {author} {\bibinfo {author} {\bibfnamefont {S.~L.}\ \bibnamefont
  {Braunstein}}\ and\ \bibinfo {author} {\bibfnamefont {P.}~\bibnamefont {van
  Loock}},\ }\href {http://link.aps.org/doi/10.1103/RevModPhys.77.513}
  {\bibfield  {journal} {\bibinfo  {journal} {Rev. Mod. Phys.}\ }\textbf
  {\bibinfo {volume} {77}},\ \bibinfo {pages} {513} (\bibinfo {year}
  {2005})}\BibitemShut {NoStop}%
\bibitem [{\citenamefont {Ferraro}\ \emph {et~al.}(2005)\citenamefont
  {Ferraro}, \citenamefont {Olivares},\ and\ \citenamefont
  {Paris}}]{Ferraro:2005ns}%
  \BibitemOpen
  \bibfield  {author} {\bibinfo {author} {\bibfnamefont {A.}~\bibnamefont
  {Ferraro}}, \bibinfo {author} {\bibfnamefont {S.}~\bibnamefont {Olivares}}, \
  and\ \bibinfo {author} {\bibfnamefont {M.~G.~A.}\ \bibnamefont {Paris}},\
  }\href@noop {} {\emph {\bibinfo {title} {Gaussian states in continuous
  variable quantum information}}}\ (\bibinfo  {publisher} {Bibliopolis},\
  \bibinfo {address} {Napoli},\ \bibinfo {year} {2005})\BibitemShut {NoStop}%
\bibitem [{\citenamefont {Cerf}\ \emph {et~al.}(2007)\citenamefont {Cerf},
  \citenamefont {Leuchs},\ and\ \citenamefont {Polzik}}]{CV2007:aa}%
  \BibitemOpen
  \bibinfo {editor} {\bibfnamefont {N.}~\bibnamefont {Cerf}}, \bibinfo {editor}
  {\bibfnamefont {G.}~\bibnamefont {Leuchs}}, \ and\ \bibinfo {editor}
  {\bibfnamefont {E.~S.}\ \bibnamefont {Polzik}},\ eds.,\ \href@noop {} {\emph
  {\bibinfo {title} {Quantum {I}nformation with {C}ontinuous {V}ariables of
  {A}toms and {L}ight}}}\ (\bibinfo  {publisher} {Imperial College Press},\
  \bibinfo {address} {London},\ \bibinfo {year} {2007})\BibitemShut {NoStop}%
\bibitem [{\citenamefont {Andersen}\ \emph {et~al.}(2010)\citenamefont
  {Andersen}, \citenamefont {Leuchs},\ and\ \citenamefont
  {Silberhorn}}]{Andersen:2010ng}%
  \BibitemOpen
  \bibfield  {author} {\bibinfo {author} {\bibfnamefont {U.~L.}\ \bibnamefont
  {Andersen}}, \bibinfo {author} {\bibfnamefont {G.}~\bibnamefont {Leuchs}}, \
  and\ \bibinfo {author} {\bibfnamefont {C.}~\bibnamefont {Silberhorn}},\
  }\href {\doibase 10.1002/lpor.200910010} {\bibfield  {journal} {\bibinfo
  {journal} {Laser Photonics Rev.}\ }\textbf {\bibinfo {volume} {4}},\ \bibinfo
  {pages} {337} (\bibinfo {year} {2010})}\BibitemShut {NoStop}%
\bibitem [{\citenamefont {Adesso}\ \emph {et~al.}(2014)\citenamefont {Adesso},
  \citenamefont {Ragy},\ and\ \citenamefont {Lee}}]{Adesso:2014pm}%
  \BibitemOpen
  \bibfield  {author} {\bibinfo {author} {\bibfnamefont {G.}~\bibnamefont
  {Adesso}}, \bibinfo {author} {\bibfnamefont {S.}~\bibnamefont {Ragy}}, \ and\
  \bibinfo {author} {\bibfnamefont {A.~R.}\ \bibnamefont {Lee}},\ }\href
  {\doibase 10.1142/S1230161214400010} {\bibfield  {journal} {\bibinfo
  {journal} {Open Syst. Inf. Dyn.}\ }\textbf {\bibinfo {volume} {21}},\
  \bibinfo {pages} {1440001} (\bibinfo {year} {2014})}\BibitemShut {NoStop}%
\bibitem [{\citenamefont {Ruppert}\ \emph {et~al.}(2014)\citenamefont
  {Ruppert}, \citenamefont {Usenko},\ and\ \citenamefont
  {Filip}}]{Ruppert:2014aa}%
  \BibitemOpen
  \bibfield  {author} {\bibinfo {author} {\bibfnamefont {L.}~\bibnamefont
  {Ruppert}}, \bibinfo {author} {\bibfnamefont {V.~C.}\ \bibnamefont {Usenko}},
  \ and\ \bibinfo {author} {\bibfnamefont {R.}~\bibnamefont {Filip}},\ }\href
  {\doibase 10.1103/PhysRevA.90.062310} {\bibfield  {journal} {\bibinfo
  {journal} {Phys. Rev. A}\ }\textbf {\bibinfo {volume} {90}},\ \bibinfo
  {pages} {062310} (\bibinfo {year} {2014})}\BibitemShut {NoStop}%
\bibitem [{\citenamefont {Ruppert}\ \emph {et~al.}(2019)\citenamefont
  {Ruppert}, \citenamefont {Peuntinger}, \citenamefont {Heim}, \citenamefont
  {Günthner}, \citenamefont {Usenko}, \citenamefont {Elser}, \citenamefont
  {Leuchs}, \citenamefont {Filip},\ and\ \citenamefont
  {Marquardt}}]{Ruppert:2019aa}%
  \BibitemOpen
  \bibfield  {author} {\bibinfo {author} {\bibfnamefont {L.}~\bibnamefont
  {Ruppert}}, \bibinfo {author} {\bibfnamefont {C.}~\bibnamefont {Peuntinger}},
  \bibinfo {author} {\bibfnamefont {B.}~\bibnamefont {Heim}}, \bibinfo {author}
  {\bibfnamefont {K.}~\bibnamefont {Günthner}}, \bibinfo {author}
  {\bibfnamefont {V.~C.}\ \bibnamefont {Usenko}}, \bibinfo {author}
  {\bibfnamefont {D.}~\bibnamefont {Elser}}, \bibinfo {author} {\bibfnamefont
  {G.}~\bibnamefont {Leuchs}}, \bibinfo {author} {\bibfnamefont
  {R.}~\bibnamefont {Filip}}, \ and\ \bibinfo {author} {\bibfnamefont
  {C.}~\bibnamefont {Marquardt}},\ }\href {\doibase 10.1088/1367-2630/ab5dd3}
  {\bibfield  {journal} {\bibinfo  {journal} {New Journal of Physics}\ }\textbf
  {\bibinfo {volume} {21}},\ \bibinfo {pages} {123036} (\bibinfo {year}
  {2019})}\BibitemShut {NoStop}%
\bibitem [{\citenamefont {Lorenz}\ \emph {et~al.}(2004)\citenamefont {Lorenz},
  \citenamefont {Korolkova},\ and\ \citenamefont {Leuchs}}]{Lorenz:2004aa}%
  \BibitemOpen
  \bibfield  {author} {\bibinfo {author} {\bibfnamefont {S.}~\bibnamefont
  {Lorenz}}, \bibinfo {author} {\bibfnamefont {N.}~\bibnamefont {Korolkova}}, \
  and\ \bibinfo {author} {\bibfnamefont {G.}~\bibnamefont {Leuchs}},\ }\href
  {http://dx.doi.org/10.1007/s00340-004-1574-7} {\bibfield  {journal} {\bibinfo
   {journal} {Appl. Phys. B}\ }\textbf {\bibinfo {volume} {79}},\ \bibinfo
  {pages} {273} (\bibinfo {year} {2004})}\BibitemShut {NoStop}%
\bibitem [{\citenamefont {Lance}\ \emph {et~al.}(2005)\citenamefont {Lance},
  \citenamefont {Symul}, \citenamefont {Sharma}, \citenamefont {Weedbrook},
  \citenamefont {Ralph},\ and\ \citenamefont {Lam}}]{Lance:2005aa}%
  \BibitemOpen
  \bibfield  {author} {\bibinfo {author} {\bibfnamefont {A.~M.}\ \bibnamefont
  {Lance}}, \bibinfo {author} {\bibfnamefont {T.}~\bibnamefont {Symul}},
  \bibinfo {author} {\bibfnamefont {V.}~\bibnamefont {Sharma}}, \bibinfo
  {author} {\bibfnamefont {C.}~\bibnamefont {Weedbrook}}, \bibinfo {author}
  {\bibfnamefont {T.~C.}\ \bibnamefont {Ralph}}, \ and\ \bibinfo {author}
  {\bibfnamefont {P.~K.}\ \bibnamefont {Lam}},\ }\href
  {http://dx.doi.org/10.1103/PhysRevLett.95.180503} {\bibfield  {journal}
  {\bibinfo  {journal} {Phys. Rev. Lett.}\ }\textbf {\bibinfo {volume} {95}},\
  \bibinfo {pages} {180503} (\bibinfo {year} {2005})}\BibitemShut {NoStop}%
\bibitem [{\citenamefont {Scarani}\ \emph {et~al.}(2009)\citenamefont
  {Scarani}, \citenamefont {Bechmann-Pasquinucci}, \citenamefont {Cerf},
  \citenamefont {Du{\v s}ek}, \citenamefont {L{\"u}tkenhaus},\ and\
  \citenamefont {Peev}}]{Scarani:2009cq}%
  \BibitemOpen
  \bibfield  {author} {\bibinfo {author} {\bibfnamefont {V.}~\bibnamefont
  {Scarani}}, \bibinfo {author} {\bibfnamefont {H.}~\bibnamefont
  {Bechmann-Pasquinucci}}, \bibinfo {author} {\bibfnamefont {N.~J.}\
  \bibnamefont {Cerf}}, \bibinfo {author} {\bibfnamefont {M.}~\bibnamefont
  {Du{\v s}ek}}, \bibinfo {author} {\bibfnamefont {N.}~\bibnamefont
  {L{\"u}tkenhaus}}, \ and\ \bibinfo {author} {\bibfnamefont {M.}~\bibnamefont
  {Peev}},\ }\href {http://dx.doi.org/10.1103/RevModPhys.81.1301} {\bibfield
  {journal} {\bibinfo  {journal} {Rev. Mod. Phys.}\ }\textbf {\bibinfo {volume}
  {81}},\ \bibinfo {pages} {1301} (\bibinfo {year} {2009})}\BibitemShut
  {NoStop}%
\bibitem [{\citenamefont {Weedbrook}\ \emph {et~al.}(2012)\citenamefont
  {Weedbrook}, \citenamefont {Pirandola}, \citenamefont
  {Garc{\'\i}a-Patr{\'o}n}, \citenamefont {Cerf}, \citenamefont {Ralph},
  \citenamefont {Shapiro},\ and\ \citenamefont {Lloyd}}]{Weedbrook:2012ag}%
  \BibitemOpen
  \bibfield  {author} {\bibinfo {author} {\bibfnamefont {C.}~\bibnamefont
  {Weedbrook}}, \bibinfo {author} {\bibfnamefont {S.}~\bibnamefont
  {Pirandola}}, \bibinfo {author} {\bibfnamefont {R.}~\bibnamefont
  {Garc{\'\i}a-Patr{\'o}n}}, \bibinfo {author} {\bibfnamefont {N.~J.}\
  \bibnamefont {Cerf}}, \bibinfo {author} {\bibfnamefont {T.~C.}\ \bibnamefont
  {Ralph}}, \bibinfo {author} {\bibfnamefont {J.~H.}\ \bibnamefont {Shapiro}},
  \ and\ \bibinfo {author} {\bibfnamefont {S.}~\bibnamefont {Lloyd}},\ }\href
  {http://link.aps.org/doi/10.1103/RevModPhys.84.621} {\bibfield  {journal}
  {\bibinfo  {journal} {Rev. Mod. Phys.}\ }\textbf {\bibinfo {volume} {84}},\
  \bibinfo {pages} {621} (\bibinfo {year} {2012})}\BibitemShut {NoStop}%
\bibitem [{\citenamefont {Wigner}(1932)}]{Wigner:1932aa}%
  \BibitemOpen
  \bibfield  {author} {\bibinfo {author} {\bibfnamefont {E.~P.}\ \bibnamefont
  {Wigner}},\ }\href {http://dx.doi.org/10.1103/PhysRev.40.749} {\bibfield
  {journal} {\bibinfo  {journal} {Phys. Rev.}\ }\textbf {\bibinfo {volume}
  {40}},\ \bibinfo {pages} {749} (\bibinfo {year} {1932})}\BibitemShut
  {NoStop}%
\bibitem [{\citenamefont {Cahill}\ and\ \citenamefont
  {Glauber}(1969)}]{Cahill:1969qd}%
  \BibitemOpen
  \bibfield  {author} {\bibinfo {author} {\bibfnamefont {K.~E.}\ \bibnamefont
  {Cahill}}\ and\ \bibinfo {author} {\bibfnamefont {R.~J.}\ \bibnamefont
  {Glauber}},\ }\href {http://dx.doi.org/10.1103/PhysRev.177.1882} {\bibfield
  {journal} {\bibinfo  {journal} {Phys. Rev.}\ }\textbf {\bibinfo {volume}
  {177}},\ \bibinfo {pages} {1882} (\bibinfo {year} {1969})}\BibitemShut
  {NoStop}%
\bibitem [{\citenamefont {Holevo}\ \emph {et~al.}(1999)\citenamefont {Holevo},
  \citenamefont {Sohma},\ and\ \citenamefont {Hirota}}]{Holevo:1999aa}%
  \BibitemOpen
  \bibfield  {author} {\bibinfo {author} {\bibfnamefont {A.~S.}\ \bibnamefont
  {Holevo}}, \bibinfo {author} {\bibfnamefont {M.}~\bibnamefont {Sohma}}, \
  and\ \bibinfo {author} {\bibfnamefont {O.}~\bibnamefont {Hirota}},\ }\href
  {\doibase 10.1103/PhysRevA.59.1820} {\bibfield  {journal} {\bibinfo
  {journal} {Phys. Rev. A}\ }\textbf {\bibinfo {volume} {59}},\ \bibinfo
  {pages} {1820} (\bibinfo {year} {1999})}\BibitemShut {NoStop}%
\bibitem [{\citenamefont {Eisert}\ and\ \citenamefont
  {Wolf}(2007)}]{Eisert:2007aa}%
  \BibitemOpen
  \bibfield  {author} {\bibinfo {author} {\bibfnamefont {J.}~\bibnamefont
  {Eisert}}\ and\ \bibinfo {author} {\bibfnamefont {M.~M.}\ \bibnamefont
  {Wolf}},\ }\href@noop {} {\emph {\bibinfo {title} {Quantum Information with
  Continuous Variables of Atoms and Light}}}\ (\bibinfo  {publisher} {Imperial
  College Press},\ \bibinfo {address} {London},\ \bibinfo {year}
  {2007})\BibitemShut {NoStop}%
\bibitem [{\citenamefont {Holevo}(2007)}]{Holevo:2007aa}%
  \BibitemOpen
  \bibfield  {author} {\bibinfo {author} {\bibfnamefont {A.~S.}\ \bibnamefont
  {Holevo}},\ }\href {https://doi.org/10.1134/S0032946007010012} {\bibfield
  {journal} {\bibinfo  {journal} {Probl. Inf. Transm.}\ }\textbf {\bibinfo
  {volume} {43}},\ \bibinfo {pages} {1} (\bibinfo {year} {2007})}\BibitemShut
  {NoStop}%
\bibitem [{\citenamefont {Smith}\ \emph {et~al.}(2011)\citenamefont {Smith},
  \citenamefont {Smolin},\ and\ \citenamefont {Yard}}]{Smith:2011aa}%
  \BibitemOpen
  \bibfield  {author} {\bibinfo {author} {\bibfnamefont {G.}~\bibnamefont
  {Smith}}, \bibinfo {author} {\bibfnamefont {J.}~\bibnamefont {Smolin}}, \
  and\ \bibinfo {author} {\bibfnamefont {J.}~\bibnamefont {Yard}},\ }\href
  {https://doi.org/10.1038/nphoton.2011.203} {\bibfield  {journal} {\bibinfo
  {journal} {Nat. Photon.}\ }\textbf {\bibinfo {volume} {5}},\ \bibinfo {pages}
  {624} (\bibinfo {year} {2011})}\BibitemShut {NoStop}%
\bibitem [{\citenamefont {Lupo}\ \emph {et~al.}(2011)\citenamefont {Lupo},
  \citenamefont {Pirandola}, \citenamefont {Aniello},\ and\ \citenamefont
  {Mancini}}]{Lupo:2011aa}%
  \BibitemOpen
  \bibfield  {author} {\bibinfo {author} {\bibfnamefont {C.}~\bibnamefont
  {Lupo}}, \bibinfo {author} {\bibfnamefont {S.}~\bibnamefont {Pirandola}},
  \bibinfo {author} {\bibfnamefont {P.}~\bibnamefont {Aniello}}, \ and\
  \bibinfo {author} {\bibfnamefont {S.}~\bibnamefont {Mancini}},\ }\href
  {\doibase 10.1088/0031-8949/2011/t143/014016} {\bibfield  {journal} {\bibinfo
   {journal} {Physica Scripta}\ }\textbf {\bibinfo {volume} {T143}},\ \bibinfo
  {pages} {014016} (\bibinfo {year} {2011})}\BibitemShut {NoStop}%
\bibitem [{\citenamefont {Holevo}\ and\ \citenamefont
  {Giovannetti}(2012)}]{Holevo:2012aa}%
  \BibitemOpen
  \bibfield  {author} {\bibinfo {author} {\bibfnamefont {A.~S.}\ \bibnamefont
  {Holevo}}\ and\ \bibinfo {author} {\bibfnamefont {V.}~\bibnamefont
  {Giovannetti}},\ }\href {\doibase 10.1088/0034-4885/75/4/046001} {\bibfield
  {journal} {\bibinfo  {journal} {Rep. Prog. Phys.}\ }\textbf {\bibinfo
  {volume} {75}},\ \bibinfo {pages} {046001} (\bibinfo {year}
  {2012})}\BibitemShut {NoStop}%
\bibitem [{\citenamefont {Siudzi\ifmmode~\acute{n}\else \'{n}\fi{}ska}\ \emph
  {et~al.}(2019)\citenamefont {Siudzi\ifmmode~\acute{n}\else \'{n}\fi{}ska},
  \citenamefont {Luoma},\ and\ \citenamefont {Strunz}}]{Siudzinska:2019aa}%
  \BibitemOpen
  \bibfield  {author} {\bibinfo {author} {\bibfnamefont {K.}~\bibnamefont
  {Siudzi\ifmmode~\acute{n}\else \'{n}\fi{}ska}}, \bibinfo {author}
  {\bibfnamefont {K.}~\bibnamefont {Luoma}}, \ and\ \bibinfo {author}
  {\bibfnamefont {W.~T.}\ \bibnamefont {Strunz}},\ }\href {\doibase
  10.1103/PhysRevA.100.062308} {\bibfield  {journal} {\bibinfo  {journal}
  {Phys. Rev. A}\ }\textbf {\bibinfo {volume} {100}},\ \bibinfo {pages}
  {062308} (\bibinfo {year} {2019})}\BibitemShut {NoStop}%
\bibitem [{\citenamefont {Braunstein}\ and\ \citenamefont
  {Caves}(1994)}]{Braunstein:1994aa}%
  \BibitemOpen
  \bibfield  {author} {\bibinfo {author} {\bibfnamefont {S.~L.}\ \bibnamefont
  {Braunstein}}\ and\ \bibinfo {author} {\bibfnamefont {C.~M.}\ \bibnamefont
  {Caves}},\ }\href {https://dx.doi.org/10.1103/PhysRevLett.72.3439} {\bibfield
   {journal} {\bibinfo  {journal} {Phys. Rev. Lett.}\ }\textbf {\bibinfo
  {volume} {72}},\ \bibinfo {pages} {3439} (\bibinfo {year}
  {1994})}\BibitemShut {NoStop}%
\bibitem [{\citenamefont {Gross}\ and\ \citenamefont
  {Caves}(2020)}]{Gross:2020aa}%
  \BibitemOpen
  \bibfield  {author} {\bibinfo {author} {\bibfnamefont {J.~A.}\ \bibnamefont
  {Gross}}\ and\ \bibinfo {author} {\bibfnamefont {C.~M.}\ \bibnamefont
  {Caves}},\ }\href
  {http://iopscience.iop.org/article/10.1088/1751-8121/abb9ed} {\bibfield
  {journal} {\bibinfo  {journal} {Journal of Physics A: Mathematical and
  Theoretical}\ } (\bibinfo {year} {2020})}\BibitemShut {NoStop}%
\bibitem [{\citenamefont {Kull}\ \emph {et~al.}(2020)\citenamefont {Kull},
  \citenamefont {Gu{\'{e}}rin},\ and\ \citenamefont
  {Verstraete}}]{Kull:2020aa}%
  \BibitemOpen
  \bibfield  {author} {\bibinfo {author} {\bibfnamefont {I.}~\bibnamefont
  {Kull}}, \bibinfo {author} {\bibfnamefont {P.~A.}\ \bibnamefont
  {Gu{\'{e}}rin}}, \ and\ \bibinfo {author} {\bibfnamefont {F.}~\bibnamefont
  {Verstraete}},\ }\href {\doibase 10.1088/1751-8121/ab7f67} {\bibfield
  {journal} {\bibinfo  {journal} {Journal of Physics A: Mathematical and
  Theoretical}\ }\textbf {\bibinfo {volume} {53}},\ \bibinfo {pages} {244001}
  (\bibinfo {year} {2020})}\BibitemShut {NoStop}%
\bibitem [{\citenamefont {Demkowicz-Dobrza{\'{n}}ski}\ \emph
  {et~al.}(2020)\citenamefont {Demkowicz-Dobrza{\'{n}}ski}, \citenamefont
  {G{\'{o}}recki},\ and\ \citenamefont {Gu{\c{t}}{\u{a}}}}]{DD:2020aa}%
  \BibitemOpen
  \bibfield  {author} {\bibinfo {author} {\bibfnamefont {R.}~\bibnamefont
  {Demkowicz-Dobrza{\'{n}}ski}}, \bibinfo {author} {\bibfnamefont
  {W.}~\bibnamefont {G{\'{o}}recki}}, \ and\ \bibinfo {author} {\bibfnamefont
  {M.}~\bibnamefont {Gu{\c{t}}{\u{a}}}},\ }\href {\doibase
  10.1088/1751-8121/ab8ef3} {\bibfield  {journal} {\bibinfo  {journal} {Journal
  of Physics A: Mathematical and Theoretical}\ }\textbf {\bibinfo {volume}
  {53}},\ \bibinfo {pages} {363001} (\bibinfo {year} {2020})}\BibitemShut
  {NoStop}%
\bibitem [{\citenamefont {\ifmmode~\check{S}\else \v{S}\fi{}afr\'anek}\ and\
  \citenamefont {Fuentes}(2016)}]{Safranek:2016aa}%
  \BibitemOpen
  \bibfield  {author} {\bibinfo {author} {\bibfnamefont {D.}~\bibnamefont
  {\ifmmode~\check{S}\else \v{S}\fi{}afr\'anek}}\ and\ \bibinfo {author}
  {\bibfnamefont {I.}~\bibnamefont {Fuentes}},\ }\href {\doibase
  10.1103/PhysRevA.94.062313} {\bibfield  {journal} {\bibinfo  {journal} {Phys.
  Rev. A}\ }\textbf {\bibinfo {volume} {94}},\ \bibinfo {pages} {062313}
  (\bibinfo {year} {2016})}\BibitemShut {NoStop}%
\bibitem [{\citenamefont {Nichols}\ \emph {et~al.}(2018)\citenamefont
  {Nichols}, \citenamefont {Liuzzo-Scorpo}, \citenamefont {Knott},\ and\
  \citenamefont {Adesso}}]{Rosanna:2018aa}%
  \BibitemOpen
  \bibfield  {author} {\bibinfo {author} {\bibfnamefont {R.}~\bibnamefont
  {Nichols}}, \bibinfo {author} {\bibfnamefont {P.}~\bibnamefont
  {Liuzzo-Scorpo}}, \bibinfo {author} {\bibfnamefont {P.~A.}\ \bibnamefont
  {Knott}}, \ and\ \bibinfo {author} {\bibfnamefont {G.}~\bibnamefont
  {Adesso}},\ }\href {\doibase 10.1103/PhysRevA.98.012114} {\bibfield
  {journal} {\bibinfo  {journal} {Phys. Rev. A}\ }\textbf {\bibinfo {volume}
  {98}},\ \bibinfo {pages} {012114} (\bibinfo {year} {2018})}\BibitemShut
  {NoStop}%
\bibitem [{\citenamefont {Oh}\ \emph {et~al.}(2019)\citenamefont {Oh},
  \citenamefont {Teo},\ and\ \citenamefont {Jeong}}]{Oh:2019aa}%
  \BibitemOpen
  \bibfield  {author} {\bibinfo {author} {\bibfnamefont {C.}~\bibnamefont
  {Oh}}, \bibinfo {author} {\bibfnamefont {Y.~S.}\ \bibnamefont {Teo}}, \ and\
  \bibinfo {author} {\bibfnamefont {H.}~\bibnamefont {Jeong}},\ }\href
  {https://dx.doi.org/10.1103/PhysRevLett.123.040602} {\bibfield  {journal}
  {\bibinfo  {journal} {Phys. Rev. Lett.}\ }\textbf {\bibinfo {volume} {123}},\
  \bibinfo {pages} {040602} (\bibinfo {year} {2019})}\BibitemShut {NoStop}%
\bibitem [{\citenamefont {Rahimi-Keshari}\ \emph {et~al.}(2011)\citenamefont
  {Rahimi-Keshari}, \citenamefont {Scherer}, \citenamefont {Mann},
  \citenamefont {Rezakhani}, \citenamefont {Lvovsky},\ and\ \citenamefont
  {Sanders}}]{Rahimi-Keshari:2011aa}%
  \BibitemOpen
  \bibfield  {author} {\bibinfo {author} {\bibfnamefont {S.}~\bibnamefont
  {Rahimi-Keshari}}, \bibinfo {author} {\bibfnamefont {A.}~\bibnamefont
  {Scherer}}, \bibinfo {author} {\bibfnamefont {A.}~\bibnamefont {Mann}},
  \bibinfo {author} {\bibfnamefont {A.~T.}\ \bibnamefont {Rezakhani}}, \bibinfo
  {author} {\bibfnamefont {A.~I.}\ \bibnamefont {Lvovsky}}, \ and\ \bibinfo
  {author} {\bibfnamefont {B.~C.}\ \bibnamefont {Sanders}},\ }\href {\doibase
  10.1088/1367-2630/13/1/013006} {\bibfield  {journal} {\bibinfo  {journal}
  {New Journal of Physics}\ }\textbf {\bibinfo {volume} {13}},\ \bibinfo
  {pages} {013006} (\bibinfo {year} {2011})}\BibitemShut {NoStop}%
\bibitem [{\citenamefont {Arthurs}\ and\ \citenamefont
  {Kelly}(1965)}]{Arthurs:1965al}%
  \BibitemOpen
  \bibfield  {author} {\bibinfo {author} {\bibfnamefont {E.}~\bibnamefont
  {Arthurs}}\ and\ \bibinfo {author} {\bibfnamefont {J.~L.}\ \bibnamefont
  {Kelly}},\ }\href {\doibase 10.1002/j.1538-7305.1965.tb01684.x} {\bibfield
  {journal} {\bibinfo  {journal} {Bell Syst. Tech. J.}\ }\textbf {\bibinfo
  {volume} {44}},\ \bibinfo {pages} {725} (\bibinfo {year} {1965})}\BibitemShut
  {NoStop}%
\bibitem [{\citenamefont {Yuen}(1982)}]{Yuen:1982hh}%
  \BibitemOpen
  \bibfield  {author} {\bibinfo {author} {\bibfnamefont {H.~P.}\ \bibnamefont
  {Yuen}},\ }\href {http://dx.doi.org/10.1016/0375-9601(82)90359-0} {\bibfield
  {journal} {\bibinfo  {journal} {Phys. Lett. A}\ }\textbf {\bibinfo {volume}
  {91}},\ \bibinfo {pages} {101} (\bibinfo {year} {1982})}\BibitemShut
  {NoStop}%
\bibitem [{\citenamefont {Arthurs}\ and\ \citenamefont
  {Goodman}(1988)}]{Arthurs:1988aa}%
  \BibitemOpen
  \bibfield  {author} {\bibinfo {author} {\bibfnamefont {E.}~\bibnamefont
  {Arthurs}}\ and\ \bibinfo {author} {\bibfnamefont {M.~S.}\ \bibnamefont
  {Goodman}},\ }\href {http://dx.doi.org/10.1103/PhysRevLett.60.2447}
  {\bibfield  {journal} {\bibinfo  {journal} {Phys. Rev. Lett.}\ }\textbf
  {\bibinfo {volume} {60}},\ \bibinfo {pages} {2447} (\bibinfo {year}
  {1988})}\BibitemShut {NoStop}%
\bibitem [{\citenamefont {Martens}\ and\ \citenamefont
  {de~Muynck}(1990)}]{Martens:1990al}%
  \BibitemOpen
  \bibfield  {author} {\bibinfo {author} {\bibfnamefont {H.}~\bibnamefont
  {Martens}}\ and\ \bibinfo {author} {\bibfnamefont {W.~M.}\ \bibnamefont
  {de~Muynck}},\ }\href {http://dx.doi.org/10.1007/BF00731707} {\bibfield
  {journal} {\bibinfo  {journal} {Found. Phys.}\ }\textbf {\bibinfo {volume}
  {20}},\ \bibinfo {pages} {357} (\bibinfo {year} {1990})}\BibitemShut
  {NoStop}%
\bibitem [{\citenamefont {Martens}\ and\ \citenamefont
  {de~Muynck}(1991)}]{Martens:1991aa}%
  \BibitemOpen
  \bibfield  {author} {\bibinfo {author} {\bibfnamefont {H.}~\bibnamefont
  {Martens}}\ and\ \bibinfo {author} {\bibfnamefont {W.~M.}\ \bibnamefont
  {de~Muynck}},\ }\href {http://dx.doi.org/10.1016/0375-9601(91)91015-6}
  {\bibfield  {journal} {\bibinfo  {journal} {Phys. Lett. A}\ }\textbf
  {\bibinfo {volume} {157}},\ \bibinfo {pages} {441} (\bibinfo {year}
  {1991})}\BibitemShut {NoStop}%
\bibitem [{\citenamefont {Raymer}(1994)}]{Raymer:1994aj}%
  \BibitemOpen
  \bibfield  {author} {\bibinfo {author} {\bibfnamefont {M.~G.}\ \bibnamefont
  {Raymer}},\ }\href {http://dx.doi.org/10.1119/1.17657} {\bibfield  {journal}
  {\bibinfo  {journal} {Am. J. Phys.}\ }\textbf {\bibinfo {volume} {62}},\
  \bibinfo {pages} {986} (\bibinfo {year} {1994})}\BibitemShut {NoStop}%
\bibitem [{\citenamefont {Trifonov}\ \emph {et~al.}(2001)\citenamefont
  {Trifonov}, \citenamefont {Bj{\"o}rk},\ and\ \citenamefont
  {S{\"o}derholm}}]{Trifonov:2001up}%
  \BibitemOpen
  \bibfield  {author} {\bibinfo {author} {\bibfnamefont {A.}~\bibnamefont
  {Trifonov}}, \bibinfo {author} {\bibfnamefont {G.}~\bibnamefont {Bj{\"o}rk}},
  \ and\ \bibinfo {author} {\bibfnamefont {J.}~\bibnamefont {S{\"o}derholm}},\
  }\href {http://dx.doi.org/10.1103/PhysRevLett.86.4423} {\bibfield  {journal}
  {\bibinfo  {journal} {Phys. Rev. Lett.}\ }\textbf {\bibinfo {volume} {86}},\
  \bibinfo {pages} {4423} (\bibinfo {year} {2001})}\BibitemShut {NoStop}%
\bibitem [{\citenamefont {Werner}(2004)}]{Werner:2004as}%
  \BibitemOpen
  \bibfield  {author} {\bibinfo {author} {\bibfnamefont {R.~F.}\ \bibnamefont
  {Werner}},\ }\href {http://dl.acm.org/citation.cfm?id=2011593.2011606}
  {\bibfield  {journal} {\bibinfo  {journal} {Quantum Info. Comput.}\ }\textbf
  {\bibinfo {volume} {4}},\ \bibinfo {pages} {546} (\bibinfo {year}
  {2004})}\BibitemShut {NoStop}%
\bibitem [{\citenamefont {{\v R}eh{\'a}{\v c}ek}\ \emph
  {et~al.}(2015)\citenamefont {{\v R}eh{\'a}{\v c}ek}, \citenamefont {Teo},
  \citenamefont {Hradil},\ and\ \citenamefont {Wallentowitz}}]{Rehacek:2015qp}%
  \BibitemOpen
  \bibfield  {author} {\bibinfo {author} {\bibfnamefont {J.}~\bibnamefont {{\v
  R}eh{\'a}{\v c}ek}}, \bibinfo {author} {\bibfnamefont {Y.~S.}\ \bibnamefont
  {Teo}}, \bibinfo {author} {\bibfnamefont {Z.}~\bibnamefont {Hradil}}, \ and\
  \bibinfo {author} {\bibfnamefont {S.}~\bibnamefont {Wallentowitz}},\ }\href
  {http://dx.doi.org/10.1038/srep12289} {\bibfield  {journal} {\bibinfo
  {journal} {Sci. Rep.}\ }\textbf {\bibinfo {volume} {5}},\ \bibinfo {pages}
  {12289} (\bibinfo {year} {2015})}\BibitemShut {NoStop}%
\bibitem [{\citenamefont {M\"{u}ller}\ \emph {et~al.}(2016)\citenamefont
  {M\"{u}ller}, \citenamefont {Peuntinger}, \citenamefont {Dirmeier},
  \citenamefont {Khan}, \citenamefont {Vogl}, \citenamefont {Marquardt},
  \citenamefont {Leuchs}, \citenamefont {S\'{a}nchez-Soto}, \citenamefont
  {Teo}, \citenamefont {Hradil},\ and\ \citenamefont
  {\v{R}eh\'{a}\v{c}ek}}]{Muller:2016da}%
  \BibitemOpen
  \bibfield  {author} {\bibinfo {author} {\bibfnamefont {C.~R.}\ \bibnamefont
  {M\"{u}ller}}, \bibinfo {author} {\bibfnamefont {C.}~\bibnamefont
  {Peuntinger}}, \bibinfo {author} {\bibfnamefont {T.}~\bibnamefont
  {Dirmeier}}, \bibinfo {author} {\bibfnamefont {I.}~\bibnamefont {Khan}},
  \bibinfo {author} {\bibfnamefont {U.}~\bibnamefont {Vogl}}, \bibinfo {author}
  {\bibfnamefont {C.}~\bibnamefont {Marquardt}}, \bibinfo {author}
  {\bibfnamefont {G.}~\bibnamefont {Leuchs}}, \bibinfo {author} {\bibfnamefont
  {L.~L.}\ \bibnamefont {S\'{a}nchez-Soto}}, \bibinfo {author} {\bibfnamefont
  {Y.~S.}\ \bibnamefont {Teo}}, \bibinfo {author} {\bibfnamefont
  {Z.}~\bibnamefont {Hradil}}, \ and\ \bibinfo {author} {\bibfnamefont
  {J.}~\bibnamefont {\v{R}eh\'{a}\v{c}ek}},\ }\href
  {http://dx.doi.org/10.1103/PhysRevLett.117.070801} {\bibfield  {journal}
  {\bibinfo  {journal} {Phys. Rev. Lett.}\ }\textbf {\bibinfo {volume} {117}},\
  \bibinfo {pages} {070801} (\bibinfo {year} {2016})}\BibitemShut {NoStop}%
\bibitem [{\citenamefont {Teo}\ \emph {et~al.}(2017)\citenamefont {Teo},
  \citenamefont {M\"{u}ller}, \citenamefont {Jeong}, \citenamefont {Hradil},
  \citenamefont {\v{R}eh\'{a}\v{c}ek},\ and\ \citenamefont
  {S\'{a}nchez-Soto}}]{Teo:2017aa}%
  \BibitemOpen
  \bibfield  {author} {\bibinfo {author} {\bibfnamefont {Y.~S.}\ \bibnamefont
  {Teo}}, \bibinfo {author} {\bibfnamefont {C.~R.}\ \bibnamefont {M\"{u}ller}},
  \bibinfo {author} {\bibfnamefont {H.}~\bibnamefont {Jeong}}, \bibinfo
  {author} {\bibfnamefont {Z.}~\bibnamefont {Hradil}}, \bibinfo {author}
  {\bibfnamefont {J.}~\bibnamefont {\v{R}eh\'{a}\v{c}ek}}, \ and\ \bibinfo
  {author} {\bibfnamefont {L.~L.}\ \bibnamefont {S\'{a}nchez-Soto}},\ }\href
  {http://dx.doi.org/10.1103/PhysRevA.95.042322} {\bibfield  {journal}
  {\bibinfo  {journal} {Phys. Rev. A}\ }\textbf {\bibinfo {volume} {95}},\
  \bibinfo {pages} {042322} (\bibinfo {year} {2017})}\BibitemShut {NoStop}%
\bibitem [{\citenamefont {Yuen}\ and\ \citenamefont
  {Chan}(1983)}]{Yuen:1983ba}%
  \BibitemOpen
  \bibfield  {author} {\bibinfo {author} {\bibfnamefont {H.~P.}\ \bibnamefont
  {Yuen}}\ and\ \bibinfo {author} {\bibfnamefont {V.~W.~S.}\ \bibnamefont
  {Chan}},\ }\href {\doibase 10.1364/OL.8.000177} {\bibfield  {journal}
  {\bibinfo  {journal} {Opt. Lett.}\ }\textbf {\bibinfo {volume} {8}},\
  \bibinfo {pages} {177} (\bibinfo {year} {1983})}\BibitemShut {NoStop}%
\bibitem [{\citenamefont {Abbas}\ \emph {et~al.}(1983)\citenamefont {Abbas},
  \citenamefont {Chan},\ and\ \citenamefont {Yee}}]{Abbas:1983ak}%
  \BibitemOpen
  \bibfield  {author} {\bibinfo {author} {\bibfnamefont {G.~L.}\ \bibnamefont
  {Abbas}}, \bibinfo {author} {\bibfnamefont {V.~W.~S.}\ \bibnamefont {Chan}},
  \ and\ \bibinfo {author} {\bibfnamefont {T.~K.}\ \bibnamefont {Yee}},\ }\href
  {\doibase 10.1364/OL.8.000419} {\bibfield  {journal} {\bibinfo  {journal}
  {Opt. Lett.}\ }\textbf {\bibinfo {volume} {8}},\ \bibinfo {pages} {419}
  (\bibinfo {year} {1983})}\BibitemShut {NoStop}%
\bibitem [{\citenamefont {Schumaker}(1984)}]{Schumaker:1984qm}%
  \BibitemOpen
  \bibfield  {author} {\bibinfo {author} {\bibfnamefont {B.~L.}\ \bibnamefont
  {Schumaker}},\ }\href {\doibase 10.1364/OL.9.000189} {\bibfield  {journal}
  {\bibinfo  {journal} {Opt. Lett.}\ }\textbf {\bibinfo {volume} {9}},\
  \bibinfo {pages} {189} (\bibinfo {year} {1984})}\BibitemShut {NoStop}%
\bibitem [{\citenamefont {Zhu}(2014)}]{Zhu:2014aa}%
  \BibitemOpen
  \bibfield  {author} {\bibinfo {author} {\bibfnamefont {H.}~\bibnamefont
  {Zhu}},\ }\href {http://dx.doi.org/10.1103/PhysRevA.90.012115} {\bibfield
  {journal} {\bibinfo  {journal} {Phys. Rev. A}\ }\textbf {\bibinfo {volume}
  {90}},\ \bibinfo {pages} {012115} (\bibinfo {year} {2014})}\BibitemShut
  {NoStop}%
\bibitem [{\citenamefont {Teo}(2015)}]{Teo:2015qs}%
  \BibitemOpen
  \bibfield  {author} {\bibinfo {author} {\bibfnamefont {Y.~S.}\ \bibnamefont
  {Teo}},\ }\href@noop {} {\emph {\bibinfo {title} {Introduction to
  {Q}uantum-{S}tate {E}stimation}}}\ (\bibinfo  {publisher} {World Scientific
  Publishing Co.},\ \bibinfo {address} {Singapore},\ \bibinfo {year}
  {2015})\BibitemShut {NoStop}%
\bibitem [{\citenamefont {Wang}\ \emph {et~al.}(2013)\citenamefont {Wang},
  \citenamefont {Yu}, \citenamefont {Hu}, \citenamefont {Miranowicz},\ and\
  \citenamefont {Nori}}]{Wang:2013aa}%
  \BibitemOpen
  \bibfield  {author} {\bibinfo {author} {\bibfnamefont {X.-B.}\ \bibnamefont
  {Wang}}, \bibinfo {author} {\bibfnamefont {Z.-W.}\ \bibnamefont {Yu}},
  \bibinfo {author} {\bibfnamefont {J.-Z.}\ \bibnamefont {Hu}}, \bibinfo
  {author} {\bibfnamefont {A.}~\bibnamefont {Miranowicz}}, \ and\ \bibinfo
  {author} {\bibfnamefont {F.}~\bibnamefont {Nori}},\ }\href {\doibase
  10.1103/PhysRevA.88.022101} {\bibfield  {journal} {\bibinfo  {journal} {Phys.
  Rev. A}\ }\textbf {\bibinfo {volume} {88}},\ \bibinfo {pages} {022101}
  (\bibinfo {year} {2013})}\BibitemShut {NoStop}%
\bibitem [{\citenamefont {Bongioanni}\ \emph {et~al.}(2010)\citenamefont
  {Bongioanni}, \citenamefont {Sansoni}, \citenamefont {Sciarrino},
  \citenamefont {Vallone},\ and\ \citenamefont {Mataloni}}]{Bongioanni:2010aa}%
  \BibitemOpen
  \bibfield  {author} {\bibinfo {author} {\bibfnamefont {I.}~\bibnamefont
  {Bongioanni}}, \bibinfo {author} {\bibfnamefont {L.}~\bibnamefont {Sansoni}},
  \bibinfo {author} {\bibfnamefont {F.}~\bibnamefont {Sciarrino}}, \bibinfo
  {author} {\bibfnamefont {G.}~\bibnamefont {Vallone}}, \ and\ \bibinfo
  {author} {\bibfnamefont {P.}~\bibnamefont {Mataloni}},\ }\href {\doibase
  10.1103/PhysRevA.82.042307} {\bibfield  {journal} {\bibinfo  {journal} {Phys.
  Rev. A}\ }\textbf {\bibinfo {volume} {82}},\ \bibinfo {pages} {042307}
  (\bibinfo {year} {2010})}\BibitemShut {NoStop}%
\bibitem [{\citenamefont {Teo}\ \emph {et~al.}(2020)\citenamefont {Teo},
  \citenamefont {Struchalin}, \citenamefont {Kovlakov}, \citenamefont {Ahn},
  \citenamefont {Jeong}, \citenamefont {Straupe}, \citenamefont {Kulik},
  \citenamefont {Leuchs},\ and\ \citenamefont {S\'anchez-Soto}}]{Teo:2020aa}%
  \BibitemOpen
  \bibfield  {author} {\bibinfo {author} {\bibfnamefont {Y.~S.}\ \bibnamefont
  {Teo}}, \bibinfo {author} {\bibfnamefont {G.~I.}\ \bibnamefont {Struchalin}},
  \bibinfo {author} {\bibfnamefont {E.~V.}\ \bibnamefont {Kovlakov}}, \bibinfo
  {author} {\bibfnamefont {D.}~\bibnamefont {Ahn}}, \bibinfo {author}
  {\bibfnamefont {H.}~\bibnamefont {Jeong}}, \bibinfo {author} {\bibfnamefont
  {S.~S.}\ \bibnamefont {Straupe}}, \bibinfo {author} {\bibfnamefont {S.~P.}\
  \bibnamefont {Kulik}}, \bibinfo {author} {\bibfnamefont {G.}~\bibnamefont
  {Leuchs}}, \ and\ \bibinfo {author} {\bibfnamefont {L.~L.}\ \bibnamefont
  {S\'anchez-Soto}},\ }\href {\doibase 10.1103/PhysRevA.101.022334} {\bibfield
  {journal} {\bibinfo  {journal} {Phys. Rev. A}\ }\textbf {\bibinfo {volume}
  {101}},\ \bibinfo {pages} {022334} (\bibinfo {year} {2020})}\BibitemShut
  {NoStop}%
\bibitem [{\citenamefont {Miyata}\ \emph {et~al.}(2014)\citenamefont {Miyata},
  \citenamefont {Ogawa}, \citenamefont {Marek}, \citenamefont {Filip},
  \citenamefont {Yonezawa}, \citenamefont {Yoshikawa},\ and\ \citenamefont
  {Furusawa}}]{PhysRevA.90.060302}%
  \BibitemOpen
  \bibfield  {author} {\bibinfo {author} {\bibfnamefont {K.}~\bibnamefont
  {Miyata}}, \bibinfo {author} {\bibfnamefont {H.}~\bibnamefont {Ogawa}},
  \bibinfo {author} {\bibfnamefont {P.}~\bibnamefont {Marek}}, \bibinfo
  {author} {\bibfnamefont {R.}~\bibnamefont {Filip}}, \bibinfo {author}
  {\bibfnamefont {H.}~\bibnamefont {Yonezawa}}, \bibinfo {author}
  {\bibfnamefont {J.-i.}\ \bibnamefont {Yoshikawa}}, \ and\ \bibinfo {author}
  {\bibfnamefont {A.}~\bibnamefont {Furusawa}},\ }\href {\doibase
  10.1103/PhysRevA.90.060302} {\bibfield  {journal} {\bibinfo  {journal} {Phys.
  Rev. A}\ }\textbf {\bibinfo {volume} {90}},\ \bibinfo {pages} {060302}
  (\bibinfo {year} {2014})}\BibitemShut {NoStop}%
\bibitem [{\citenamefont {Kashiwazaki}\ \emph {et~al.}(2020)\citenamefont
  {Kashiwazaki}, \citenamefont {Takanashi}, \citenamefont {Yamashima},
  \citenamefont {Kazama}, \citenamefont {Enbutsu}, \citenamefont {Kasahara},
  \citenamefont {Umeki},\ and\ \citenamefont
  {Furusawa}}]{APLPhotonics.5.036104}%
  \BibitemOpen
  \bibfield  {author} {\bibinfo {author} {\bibfnamefont {T.}~\bibnamefont
  {Kashiwazaki}}, \bibinfo {author} {\bibfnamefont {N.}~\bibnamefont
  {Takanashi}}, \bibinfo {author} {\bibfnamefont {T.}~\bibnamefont
  {Yamashima}}, \bibinfo {author} {\bibfnamefont {T.}~\bibnamefont {Kazama}},
  \bibinfo {author} {\bibfnamefont {K.}~\bibnamefont {Enbutsu}}, \bibinfo
  {author} {\bibfnamefont {R.}~\bibnamefont {Kasahara}}, \bibinfo {author}
  {\bibfnamefont {T.}~\bibnamefont {Umeki}}, \ and\ \bibinfo {author}
  {\bibfnamefont {A.}~\bibnamefont {Furusawa}},\ }\href {\doibase
  10.1063/1.5142437} {\bibfield  {journal} {\bibinfo  {journal} {APL
  Photonics}\ }\textbf {\bibinfo {volume} {5}},\ \bibinfo {pages} {036104}
  (\bibinfo {year} {2020})}\BibitemShut {NoStop}%
\bibitem [{\citenamefont {Takeda}\ and\ \citenamefont
  {Furusawa}(2019)}]{APLPhotonics.4.060902}%
  \BibitemOpen
  \bibfield  {author} {\bibinfo {author} {\bibfnamefont {S.}~\bibnamefont
  {Takeda}}\ and\ \bibinfo {author} {\bibfnamefont {A.}~\bibnamefont
  {Furusawa}},\ }\href {\doibase 10.1063/1.5100160} {\bibfield  {journal}
  {\bibinfo  {journal} {APL Photonics}\ }\textbf {\bibinfo {volume} {4}},\
  \bibinfo {pages} {060902} (\bibinfo {year} {2019})}\BibitemShut {NoStop}%
\bibitem [{\citenamefont {Josse}\ \emph {et~al.}(2006)\citenamefont {Josse},
  \citenamefont {Sabuncu}, \citenamefont {Cerf}, \citenamefont {Leuchs},\ and\
  \citenamefont {Andersen}}]{PhysRevLett.96.163602}%
  \BibitemOpen
  \bibfield  {author} {\bibinfo {author} {\bibfnamefont {V.}~\bibnamefont
  {Josse}}, \bibinfo {author} {\bibfnamefont {M.}~\bibnamefont {Sabuncu}},
  \bibinfo {author} {\bibfnamefont {N.~J.}\ \bibnamefont {Cerf}}, \bibinfo
  {author} {\bibfnamefont {G.}~\bibnamefont {Leuchs}}, \ and\ \bibinfo {author}
  {\bibfnamefont {U.~L.}\ \bibnamefont {Andersen}},\ }\href {\doibase
  10.1103/PhysRevLett.96.163602} {\bibfield  {journal} {\bibinfo  {journal}
  {Phys. Rev. Lett.}\ }\textbf {\bibinfo {volume} {96}},\ \bibinfo {pages}
  {163602} (\bibinfo {year} {2006})}\BibitemShut {NoStop}%
\bibitem [{\citenamefont {Bowen}\ \emph {et~al.}(2003)\citenamefont {Bowen},
  \citenamefont {Treps}, \citenamefont {Buchler}, \citenamefont {Schnabel},
  \citenamefont {Ralph}, \citenamefont {Symul},\ and\ \citenamefont
  {Lam}}]{IEEEquantumelectronics.9.1519}%
  \BibitemOpen
  \bibfield  {author} {\bibinfo {author} {\bibfnamefont {W.~P.}\ \bibnamefont
  {Bowen}}, \bibinfo {author} {\bibfnamefont {N.}~\bibnamefont {Treps}},
  \bibinfo {author} {\bibfnamefont {B.~C.}\ \bibnamefont {Buchler}}, \bibinfo
  {author} {\bibfnamefont {R.}~\bibnamefont {Schnabel}}, \bibinfo {author}
  {\bibfnamefont {T.~C.}\ \bibnamefont {Ralph}}, \bibinfo {author}
  {\bibfnamefont {T.}~\bibnamefont {Symul}}, \ and\ \bibinfo {author}
  {\bibfnamefont {P.~K.}\ \bibnamefont {Lam}},\ }\href@noop {} {\bibfield
  {journal} {\bibinfo  {journal} {IEEE Journal of selected topics in quantum
  electronics}\ }\textbf {\bibinfo {volume} {9}},\ \bibinfo {pages} {1519}
  (\bibinfo {year} {2003})}\BibitemShut {NoStop}%
\bibitem [{\citenamefont {Filippov}\ and\ \citenamefont
  {Ziman}(2014)}]{PhysRevA.90.010301}%
  \BibitemOpen
  \bibfield  {author} {\bibinfo {author} {\bibfnamefont {S.~N.}\ \bibnamefont
  {Filippov}}\ and\ \bibinfo {author} {\bibfnamefont {M.}~\bibnamefont
  {Ziman}},\ }\href@noop {} {\bibfield  {journal} {\bibinfo  {journal} {Phys.
  Rev. A}\ }\textbf {\bibinfo {volume} {90}},\ \bibinfo {pages} {010301}
  (\bibinfo {year} {2014})}\BibitemShut {NoStop}%
\bibitem [{\citenamefont {Pe{\v r}ina}(1984)}]{Perina:2001aa}%
  \BibitemOpen
  \bibfield  {author} {\bibinfo {author} {\bibfnamefont {J.}~\bibnamefont
  {Pe{\v r}ina}},\ }\href@noop {} {\emph {\bibinfo {title} {Quantum Statistics
  of Linear and Nonlinear Optical Phenomena}}}\ (\bibinfo  {publisher}
  {Springer},\ \bibinfo {address} {Netherlands},\ \bibinfo {year}
  {1984})\BibitemShut {NoStop}%
\bibitem [{\citenamefont {Agrawal}\ and\ \citenamefont
  {Mehta}(1977)}]{Agrawal:1977aa}%
  \BibitemOpen
  \bibfield  {author} {\bibinfo {author} {\bibfnamefont {G.~P.}\ \bibnamefont
  {Agrawal}}\ and\ \bibinfo {author} {\bibfnamefont {C.~L.}\ \bibnamefont
  {Mehta}},\ }\href {https://doi.org/10.1063/1.523283} {\bibfield  {journal}
  {\bibinfo  {journal} {J. Math. Phys}\ }\textbf {\bibinfo {volume} {18}},\
  \bibinfo {pages} {408} (\bibinfo {year} {1977})}\BibitemShut {NoStop}%
\end{thebibliography}
